\documentclass[review]{fcs}

\usepackage{diagbox}
\usepackage{multirow}
\usepackage{algorithm}
\usepackage{algpseudocode}
\usepackage{enumitem}
\usepackage{hyperref}
\usepackage{breakurl} % Helps in breaking long URLs
\usepackage{graphicx}
\usepackage{svg}
\usepackage{microtype}
\usepackage[hyphens]{xurl}
\usepackage{tablefootnote}
\usepackage{longtable}
\usepackage{threeparttable}
\usepackage{xcolor}
\usepackage{hyperref}

\hypersetup{
	breaklinks=true
}

\title{A Survey on Sequential Recommendation}
\shorttitle{}
\author[1]{Liwei Pan}
\author*[1]{Weike Pan}
\author[1]{Meiyan Wei}
\author[2]{Hongzhi Yin}
\author[3]{Zhong Ming}
\address[1]{College of Computer Science and Software Engineering, Shenzhen University, Shenzhen 518060, China}
\address[2]{School of Electrical Engineering and Computer Science, The University of Queensland, Queensland 4072, Australia}
\address[3]{College of Big Data and Internet, Shenzhen Technology University, Shenzhen 518118, China}
\fcssetup{
	received       = {12 07, 2024},
	accepted       = {month dd, yyyy},
	corr-email     = {panweike@szu.edu.cn},
}
\begin{abstract}
	Different from most conventional recommendation problems, sequential recommendation focuses on learning users' preferences by exploiting the internal order and dependency among the interacted items, which has received significant attention from both researchers and practitioners. In recent years, we have witnessed great progress and achievements in this field, necessitating a new survey. In this survey, we study the SR problem from a new perspective (i.e., the construction of an item's properties), and summarize the most recent techniques used in sequential recommendation such as multi-modal SR, generative SR, LLM-powered SR, ultra-long SR and data-augmented SR. Moreover, we introduce some frontier research topics in sequential recommendation, e.g., open-domain SR, data-centric SR, cloud-edge collaborative SR, continuous SR, SR for good, and explainable SR. We believe that our survey could be served as a valuable roadmap for readers in this field.
\end{abstract}

\begin{document}
	
	\section{Introduction}
	
	Recommender systems are often designed to deliver items that users are interested in from a large collection, which are expected to address the information overload problem and save time for users. Meanwhile, recommender systems are usually deemed effective in improving business profits by motivating people to purchase  their items of interest. So far, they have been deployed in various real-world applications, e.g., e-commerce (e.g., Amazon and Alibaba), streaming services (e.g., YouTube and TikTok), social media (e.g., WeChat and Twitter), online advertising, and so on.

\begin{figure}[h]
		\includegraphics[width=\linewidth]{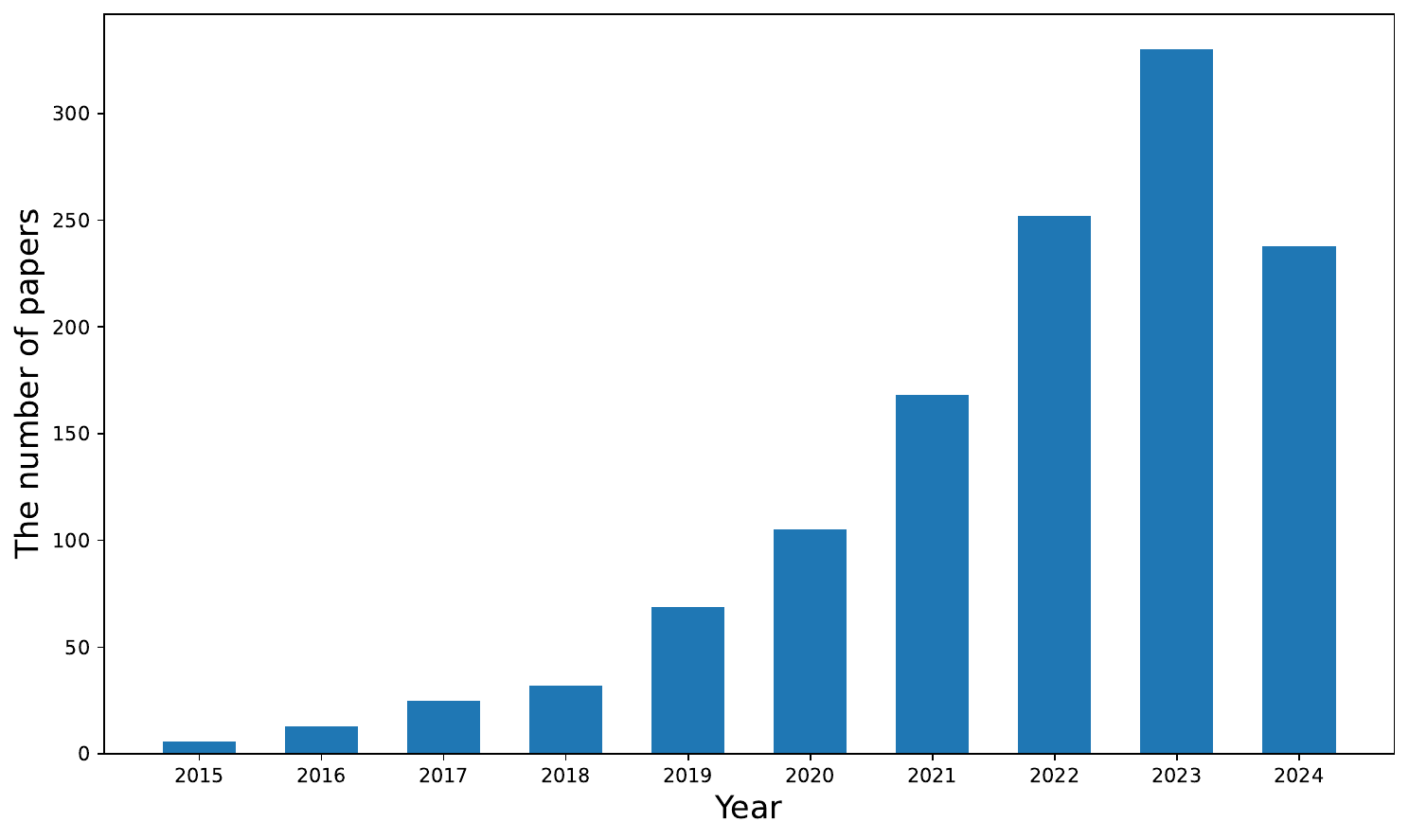}
		\caption{The number of articles on sequential recommendation published between 2015 and 2024.}
		\label{fig:numpaper}
	\end{figure}

	Various sequential recommendation (SR) models have been proposed, which have achieved significant recommendation improvements in performance recently \cite{SASRec}. The main idea of an SR model is to exploit the position and sequential information of the interacted items, so as to capture the dependencies among them. Given a user’s interaction sequence, a typical SR model aims to predict the next likely-to-be-preferred items for this user. In recent years, there has been a significant increase in the number of SR models. By searching the keywords (i.e., sequential, recommend) in the DBLP website\footnote{\url{https://dblp.org/}}, we have the specific publication numbers of papers about sequential recommendation, which are shown in Figure~\ref{fig:numpaper}. We can see that the number of papers about sequential recommendation increases with time going by. Note that the number of papers published in 2024 is slightly fewer than in 2023, because we collected the data on June 12th, 2024.
	
	Some early SR models \cite{TransRec_t, SASRec} use a unique ID to denote an item. These SR models are simple yet effective. For each item, an SR model usually needs to learn a unique item embedding, which is often efficient. Meanwhile, the item embeddings can capture the correlations among the items and extract their latent characteristics. However, these SR models have some weaknesses. For instance, only using item IDs cannot solve cold-start and sparsity problems. Therefore, some SR models \cite{ICAI-SR, NOVA} combine  item IDs and item features. The item features mainly include categorical features (e.g., categories), numerical features (e.g., prices) and graph structure features (e.g., social networks and knowledge graph).  Due to privacy issues, we often know little about the users' features. By leveraging these features, we can effectively learn item representations even when interactions with some items are sparse. Therefore, these SR models can alleviate the cold-start and data sparsity problems. However, these SR models also have some limitations. For example, they usually heavily rely on item IDs during training. If data are different, the item IDs are different. Therefore, these SR models trained on one data cannot be easily transferred to another data through fine-tuning or other methods. With the fast development of natural language processing and computer vision, models like BERT \cite{BERT} and ResNet \cite{ResNet} can extract rich features from text descriptions and images, respectively. To solve  the issue of data dependency and high-difficulty in cross-data transfer, some SR models \cite{MoRec, UniSRec} leverage BERT and ResNet to extract semantic representations of items from text descriptions and images, respectively, and then directly use them to represent the items. Due to the widespread availability of multi-modal features, these SR models trained on one data from a certain platform can be deployed on another platform through fine-tuning. However, the semantic representations extracted from multi-modal features cannot fully capture users' fine-grained preferences and collaborative signals. Therefore, some SR models turn to combine the multi-modal features with item IDs to learn users' preferences more comprehensively.
	
	LLMs are trained using a large amount of data and demonstrate promising performance. More importantly, LLMs have a strong reasoning ability. Therefore, LLMs have been applied to many applications, including recommender systems. For example, some SR models leverage LLMs to recommend items to users directly \cite{GenRec}. Some SR models leverage the output representations of LLMs for recommending personalized items more accurately since these representations contain rich semantic knowledge \cite{LLMs_based_SR}. To alleviate data sparsity and cold-start problems, some SR models leverage LLMs to generate some data. These SR models then combine the generated data with the original data to achieve better recommendation performance. However, LLMs-based SR models also have some shortcomings. For example, training LLMs-based SR models requires significant computational resources. This survey summarizes how to leverage LLMs in SR models.
	
	In sequential recommendation, modeling ultra-long interaction sequences becomes increasingly complex. Over time, users interact with more and more items. With the length of interaction sequences increasing, the time of both training and inference will increase. Meanwhile, as users interact with an increasing number of items, noise also becomes more serious in the interaction sequences. Therefore, modeling ultra-long interaction sequences becomes increasingly difficult. Common approaches retrieve a few interacted items from ultra-long interaction sequences \cite{SIM}. Our survey will summarize some common methods applied to modeling ultra-long interaction sequences modeling.
	
	In our survey, we also introduce data augmentation methods in SR models. This is because data augmentation methods can be leveraged by different SR models. They mainly focus on generating more interaction data to alleviate data sparsity and cold-start problems. Common augmentation methods include some operations, such as resorting, masking, cropping, and others on interaction sequences \cite{CL4SRec}.
	
	The contributions of our survey are summarized as follows:
	\begin{itemize}[left=0pt]
		\item Our survey summarizes the early and recent works about SR models from a more comprehensive manner.
		\item Our survey categorizes the existing SR models into four categories according to the construction of item properties.
		\item Our survey summarizes the latest techniques applied to sequential recommendation.
		\item Our survey introduces empirical studies and future research directions in SR models.
	\end{itemize}

	The remainder of our survey is organized as follows. In Section \hyperref[sec:related_surveys]{2}, we discuss some surveys on sequential recommendation and session-based recommendation. In Section \hyperref[sec:sequential_recommendation]{3}, we provide a comprehensive view about sequential recommendation, including problem definition, properties, four different kinds of SR models and challenges. In Section \hyperref[sec:pure_ID_based_SR]{4}, we discuss some pure ID-based SR models. In Section \hyperref[sec:SR_with_side_information]{5}, we present some SR models combining IDs and side information. In Section \hyperref[sec:recent_SR_advancements]{6}, we elaborate on recent advancements in sequential recommendation, including multi-modal SR, generative SR, LLM-powered SR, ultra-long sequence modeling in SR, and data augmentation in SR. In Section \hyperref[sec:empirical_studies]{7}, we introduce datasets, evaluation protocols and metrics, as well as experimental results in sequential recommendation. Finally, we provide some insights into prospects and future directions about sequential recommendation in Section \hyperref[sec:prospects_and_future_directions]{8} and draw a conclusion in Section \hyperref[sec:conclusions]{9}.
	\section{Related Surveys}\label{sec:related_surveys}
	In recent years, an increasing number of surveys on sequential and session-based recommender systems have been published \cite{Wang.al., li.al., quadrana.al., nasir.al., fang.al., Boka.al.}. They introduce some common algorithms and models for these systems. However, with the development of LLMs and other techniques, these surveys are out-of-date to some extent as LLMs have strong deduction abilities and can fully extract semantic information for recommendation. Meanwhile, some SR models introduced in these surveys still have weaknesses. For example, only using an item ID to denote an item might influence the transferability of SR models. Some surveys \cite{chen.al., chen_shu.al.} focus on cross-domain sequential recommender systems, or multi-behavior sequential recommender systems. They only introduce a small fraction of models in sequential recommendation. Compared with these surveys, our survey introduces the development of sequential recommender systems more comprehensively. There are also some surveys about recommender systems. Jing et al. \cite{jing.al.} introduce contrastive learning in recommender systems comprehensively. Lai et al. \cite{lai.al.} discuss the importance of data and some methods of data augmentation. Zhang et al. \cite{zhang.al.1} explore the usage of different kinds of side information in recommender systems. However, these surveys only focus on one specific technique in recommender systems. Compared with these surveys, we not only classify existing research on SR but also introduce a new taxonomy to comprehend,  structure and analyze these works.
	\begin{table*}[!ht]
		\caption{Some notations used in the paper.}
		\label{tab:notations}
		\centering
		\begin{tabular}{ll}
			\hline
			Notation & Description\\
			\hline
			$\mathcal{U} = \{u_{1}, u_{2}, \ldots, u_{|U|}\}$ & A set of users\\
			$\mathcal{I} = \{i_{1}, i_{2}, \ldots, i_{|I|}\}$ & A set of items\\
			$L$ & The length of an interaction sequence\\
			$S_{u} = \{s_{u,1}, s_{u,2}, \ldots, s_{u,L}\}$ & The interaction sequence of user $u$\\
			$F_{u}^{(k)} = \{f_{u,1}^{(k)}, f_{u,2}^{(k)}, \ldots, f_{u,L}^{(k)}\}$ & The $k$-th multi-modal feature w.r.t. user $u$'s interaction sequence.\\
			$N_{u}^{(k)} = \{n_{u,1}^{(k)}, n_{u,2}^{(k)}, \ldots, n_{u,L}^{(k)}\}$ & The $k$-th categorical and numerical feature w.r.t. user $u$'s interaction sequence.\\
			$K_{m}$ & The number of multi-modal features\\
			$K_{g}$ & The number of categorical and numerical features\\
			$t_{u}$ & The item that user $u$ interacts with at the most recent time step\\
			$R_{u,i}$ & The rank generated by an SR model of candidate item $i$ for user $u$\\
			$P(i|S_{u})$ & The probability in which user $u$ interacted item $i$ given historical sequence $S_{u}$\\
			\hline
			GNNs($\cdot)$, CNNs($\cdot$), Transformer($\cdot$) & The GNNs, CNNs and Transformer models\\
			Embed($\cdot$) & The item embedding layer\\
			 $\mathbf{P}_{u}^{CNNs}$, $\mathbf{P}_{u}^{GNNs}$ & The user $u$'s preferences extracted from CNNs and GNNs\\
			$\mathbf{P}_{u}^{Tr}$ & The user $u$'s preferences extracted from Transformer\\
			$\mathbf{P}_{u}^{final}$ & The user $u$'s final preferences\\
			$\mathcal{G}$ & The graph constructed by all users' interaction sequences\\
			Fuse($\cdot$) & The fusion module\\
			\hline
			BM$_{k}$($\cdot$) & The $k$-th backbone model\\
			$\mathbf{E}_{u}$ & The item embedding matrix for user $u$\\
			$\mathbf{A}_{u}^{(k)}$ &  The $k$-th feature embedding matrix for user $u$\\
			$S_{u,1}^{(k)},S_{u,2}^{(k)},\ldots,S_{u,m}^{(k)}$ & The $m$ sub-sequences divided by the $k$-th feature\\
			$\mathbf{P}_{u}^{feat}$ & the $u$'s preferences extracted from the features of interacted items\\
			\hline
			Text$_{i}$, Img$_{i}$, Vid$_{i}$ & The text descriptions, images and videos of item $i$\\
			TextEncoder(), ImageEncoder(), VideoEncoder($\cdot$) & The text encoder, image encoder and video encoder\\
			$\mathbf{e}_{i}$ &  The representations about item $i$\\
			\hline
			$\mathbf{Y}_{i} = \{y_{i,1},y_{i,2},\ldots,y_{i,l}\}$ & The $l$ tokens representing item $i$\\
			\hline
			$x_{prompt}$ & The prompt which is fed into LLMs\\
			LLMs($\cdot$) & The large language models\\
			$\hat{x}_{prompt}$ & The outputs of LLMs while inputting $x_{prompt}$\\
			\hline
			$\mathbf{P}_{u}^{short}$,$\mathbf{P}_{u}^{long}$ & The user $u$'s short-term and long-term interests\\
			Att($\cdot$),Ret($\cdot$) &  The attention and retrieval modules\\
			$S_{u,recent}$ &  The user $u$'s recently behavior sequence\\
			\hline
		\end{tabular}
	\end{table*}
	\section{Sequential Recommendation}\label{sec:sequential_recommendation}
	In this section, we will begin by defining the problem of sequential recommendation. Then, we will introduce properties about users, items, behaviors and sequences as well as different kinds of features. Next, we will describe four different types of SR models. Finally, we will discuss some challenges faced by SR models.
	\begin{figure*}[h]
		\includegraphics[width=\linewidth]{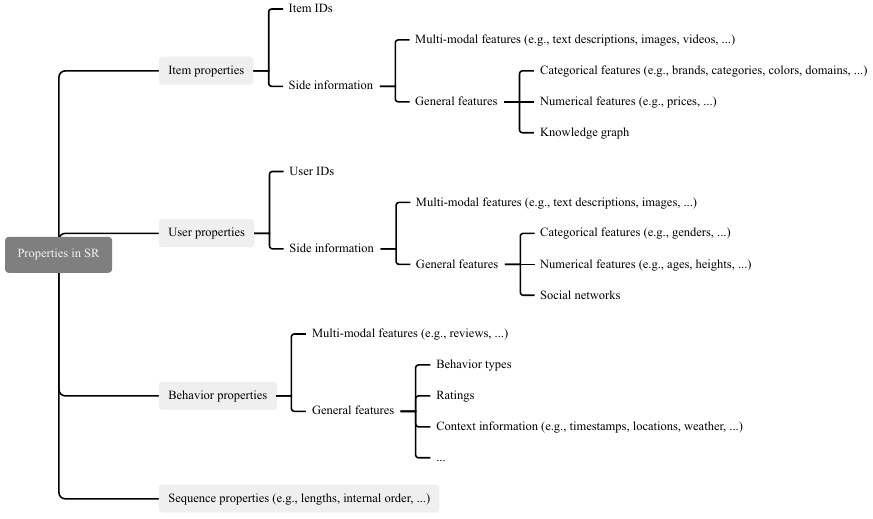}
		\caption{Properties in sequential recommendation.}
		\label{fig:SR-properties}
	\end{figure*}
	\begin{figure}[h]
		\includegraphics[width=\linewidth]{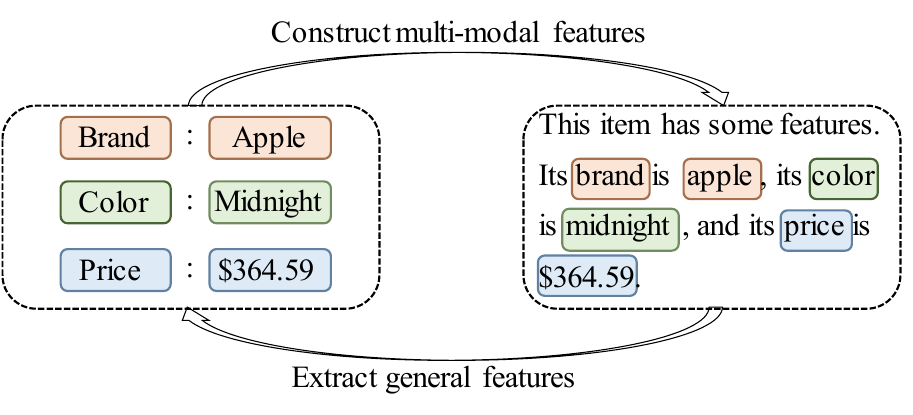}
		\caption{The conversion between multi-modal features and general features.}
		\label{fig:features-conversions}
	\end{figure}
	\subsection{Problem Definition}
	\begin{figure*}[h]
		\includegraphics[width=\linewidth]{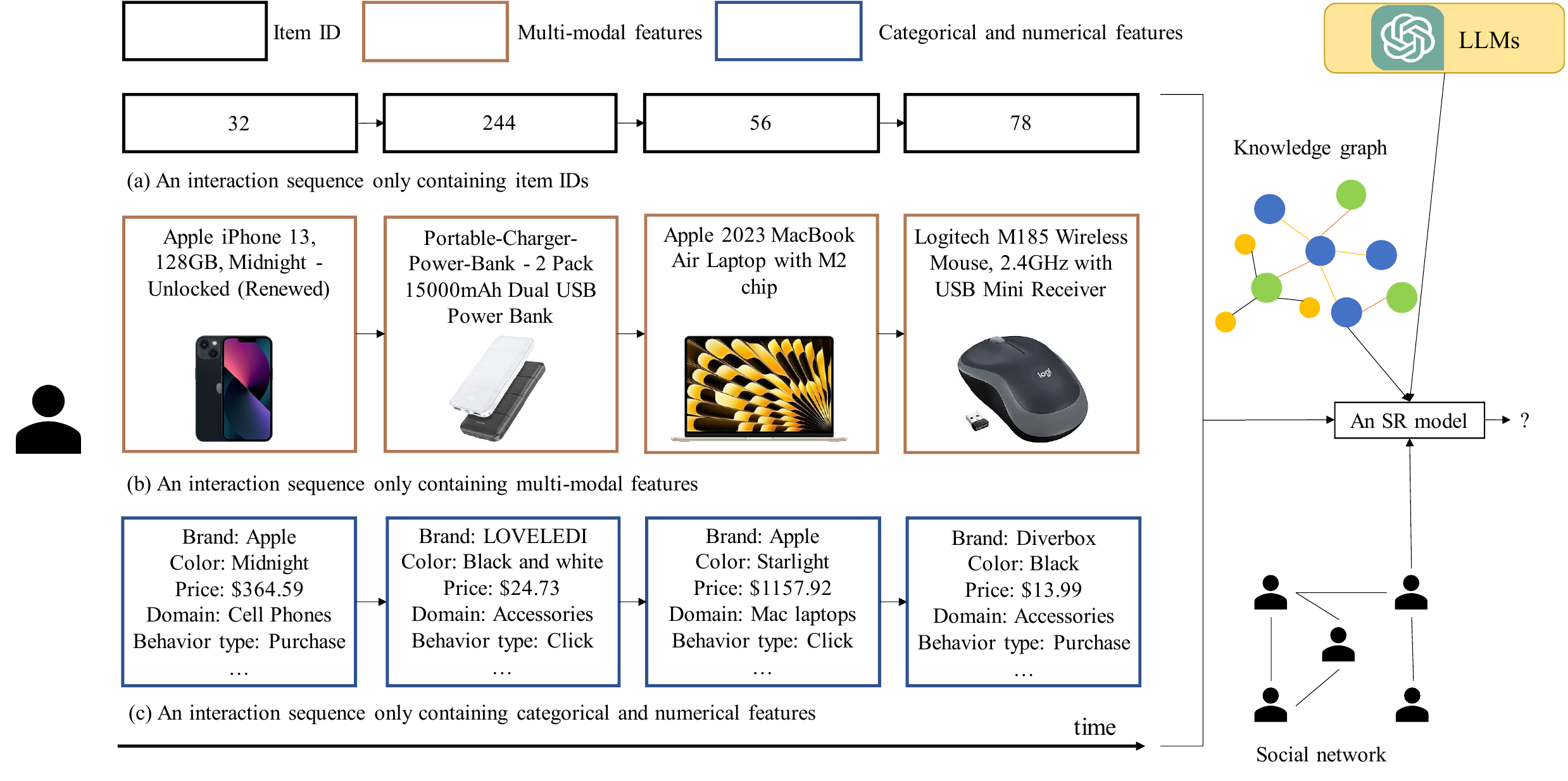}
		\caption{An illustration of sequential recommendation.}
		\label{fig:all-diagrams}
	\end{figure*}
	In sequential recommendation, we often have one or more sequences of interacted items w.r.t. each user, as well as some auxiliary information to help learn user preferences. Our goal is then to generate a ranked list of items accurately for each user.
	\subsection{Properties}
	In sequential recommendation systems, there are two important entities (i.e., an item and a user). Therefore, item properties and user properties are vital for SR models. A user interacts with items, establishing connections between the user and the items via different types of behaviors. The user's interaction sequence is constructed according to interaction order. Therefore, the properties of both the behaviors and sequences are also important in sequential recommendation systems. The specific details about these properties in sequential recommendation systems are shown in Figure~\ref{fig:SR-properties}.
	
	For item properties in SR, each item is denoted by a unique number (i.e., an item ID). Apart from the item ID, each item also has side information. The side information mainly has two types, including multi-modal features and general features. The general features have three types: categorical features, numerical features, and knowledge graph. The categorical features can be denoted by one-hot vectors. The numerical features are values represented by numbers, which have real meaning. Each numerical feature usually lies in a continuous or discrete range. User properties consist of two parts: a unique user ID and some side information specific to each user. Some specific details are shown in Figure~\ref{fig:SR-properties}. To avoid the leakage of users' privacy, the side information about users is usually unseen. Behavior properties are also comprised of general features and multi-modal features. Some features such as ratings and reviews, which might not be present in certain interactions, because not every user can provide a review or rating for an item they interact with. However, if a user interacts with an item, the behavior type of this interaction is bound to exist. Context information refers to an environment in which a behavior happens, such as a specific time and location. Sequential properties primarily include the length of interaction sequences and the internal order of items within these sequences. Meanwhile, there exist two important graph-structure features (i.e., knowledge graph and social networks).
	
	Entity IDs (i.e., item IDs and user IDs), general features and multi-modal features are quietly different in sequential recommender systems. The entity IDs are specific to a particular data. In different data or different platforms, the encoding methods for the entity IDs are quite different. For example, a movie might be encoded as item $1$ in MovieLens \footnote{\burl{https://grouplens.org/datasets/movielens/}}. However, a book might also be encoded as item 1 in the book domain of Amazon \footnote{\burl{https://huggingface.co/datasets/McAuley-Lab/Amazon-Reviews-2023}}. Therefore, item embeddings based on item IDs cannot be transferred from one platform to another platform. The general features share some common characteristics across different datasets or platforms. For example, basic behavior types in different datasets might be the same. All e-commerce platforms contain basic behavior types (i.e., click and purchase). However, the categories of items might differ across different domains or platforms, leading to different distributions of general features. If we apply the embeddings of general features in one domain to another domain, it might result in negative transfer. Techniques like a cross-attention mechanism can address this issue to some extent. Multi-modal features, such as text, images, and videos, are universal across all domains and platforms. We use language to record information as text, cameras to capture beautiful images, videos to document moments. The text, images, and videos have separate embedding spaces. Meanwhile, they all have shared space. Therefore, the multi-features can be leveraged in all domains and all platforms. If we leverage the text descriptions of an item to represent the item in a model and train the model on one platform, it can be directly applied to another platform or fine-tuned in another platform as needed. Therefore, leveraging multi-modal features to denote an item can fully realize the transfer of SR models from one platform to another.
	
	In the meantime, the general features and multi-modal features can be converted into each other. As shown in Figure~\ref{fig:features-conversions}, we can convert an item's brand, price and color into text descriptions. Then, SR models can leverage these text descriptions of this item to learn the universal representations of this item. However, this type of conversion has shortcomings. For example, if we convert timestamps to text, fine-grained time information might be damaged. Meanwhile, as shown in Figure~\ref{fig:features-conversions}, we can also extract general features from the text descriptions of an item, such as its brand, color and price. However, the extracted general features might contain some noise.
	
	There are several important features in sequential recommender systems. Different features have various meanings and play different roles in sequential recommendation. Some specific details are shown as follows:
	\begin{itemize}
		\item Timestamp refers to the precise time when a user interacts with an item. Leveraging timestamp information can capture the user’s evolving interests and periodic behaviors precisely.
		\item Category refers to the hierarchical classification of an item. Leveraging the category of the item can capture users’ preferences more precisely and narrow the range of candidate items.
		\item Price refers to the amount of money spent to purchase an item. It plays a dominant role in deciding whether to purchase the item.
		\item Text description refers to the description of an item and the title of the item. It can describe the item from different perspectives.
		\item Image refers to the visual representations of an item. It can provide a clear and intuitive understanding of the item. By looking at images, users can know what the item looks like.
		\item Rating refers to the degree of a user’s satisfaction with an item. It can reflect the user’s preferences for the item and the quality of the item.
		\item Review refers to comments that a user provides for an item. It can offer more detailed information about the user’s preferences for the item, and it can reflect the overall quality of this item.
		\item Behavior type refers to a specific behavior type in which a user interacts with an item, such as click, favorite and purchase. Leveraging different types of behaviors can capture the user's different preferences from different perspectives.
	\end{itemize}
	\subsection{Categories}
	In our surveys, we classify SR models into four categories based on the properties of items. As depicted in Figure~\ref{fig:all-diagrams}, for user $u$, there are three different interaction sequences containing item IDs, multi-modal features, and categorical and numerical features, respectively.
	
	As shown in part (a) of Figure~\ref{fig:all-diagrams}, some early works leverage a unique item ID to denote an item for recommendation \cite{SASRec, GRU4Rec}. These SR models leverage only item ID interaction sequences, which can be formalized as: $P(i|S_{u})$. In these works, given a candidate item $i$ and a user $u$'s historical interaction sequence $S_{u}$, these models need to calculate the probability $P(i|S_{u})$ in which user $u$ interacts with the candidate item $i$. We will introduce this type of SR models in Section \hyperref[sec:pure_ID_based_SR]{4}.
	
	However, SR models based on pure item IDs cannot tackle the commonly encountered cold-start and data-sparsity problems well. General features are important for recommending personalized items, which can alleviate the cold-start and data-sparsity problems and fully extract users' preferences. As depicted in part (a) and part (c) of Figure~\ref{fig:all-diagrams}, some works leverage both item ID sequences and interaction sequences containing general features (i.e., categorical features and numerical features) to recommend personalized items to users \cite{DIF-SR, NOVA}. These SR models can be formalized as: $P(i|S_{u}, N_{u}^{(1)}, N_{u}^{(2)}, \ldots, N_{u}^{(K_{g})})$. Given a user $u$'s interaction sequence $S_{u}$, which contains item IDs and general features $N_{u}^{(1)}, \ldots, N_{u}^{(K_{g})}$ embedded in the sequence, these models aim to predict the probability in which the user $u$ interacted with a candidate item $i$. To account for the influence of users' friends, social networks might be leveraged in these works \cite{STEN}. The general features of interacted items can be represented as knowledge graph. Therefore, these works might leverage knowledge graph to capture relationships among general features. We will introduce this type of SR models in Section \hyperref[sec:SR_with_side_information]{5}.
	
	Though the above-mentioned models achieve significant performance, they still have certain shortcomings. Firstly, these models heavily rely on item IDs, which hinders their transferability. Secondly, these general features might not be fully utilized by some of latest techniques effectively, such as LLMs. Therefore, some recently proposed models leverage the multi-modal features of an item to denote the item directly \cite{UniSRec, VQ-Rec}. These models can realize the transferability across different platforms. By leveraging latest techniques like LLMs, they can fully extract some semantic knowledge. These models can be formalized as: $P(i|F_{u}^{(1)}, F_{u}^{(2)}, \ldots, F_{u}^{(K_{m})})$. Given the multi-modal features derived from the items within a user $u$'s interaction sequence, these models can calculate the probability in which user $u$ interacts with a candidate item $i$. The core ideas of these models are illustrated in part (b) of Figure~\ref{fig:all-diagrams}. When both sequence information and multi-modal features (i.e., text descriptions and images) of each interaction are available, an SR model predicts the next interacted item interacted with by user $u$. By replacing item IDs with text descriptions and images, SR models can acquire rich semantic knowledge and enhance their transferability.
	
	However, relying solely on items' multi-modal features cannot capture users' fine-grained interests. In a warm-start situation, only leveraging item IDs can achieve better performance than only leveraging multi-modal features \cite{MoRec}. Therefore, some SR models leverage both item IDs and multi-modal features of these items. The simple diagram of these models is shown by combining part (a) and part (b) in Figure~\ref{fig:all-diagrams}. In these SR models, each item property contains both a unique item ID and multi-modal features. To extract semantic information from multi-modal features, LLMs can be leveraged. These SR models can be formalized as: $P(i|S_{u}, F_{u}^{(1)}, F_{u}^{(2)}, \ldots, F_{u}^{(K_{m})})$. We will introduce the aforementioned two types of SR models in Section \hyperref[sec:multi_modal_SR]{6.1} of Multi-Modal SR.
	\subsection{Taxonomy}
\begin{figure*}[!ht]
		\includegraphics[width=\linewidth]{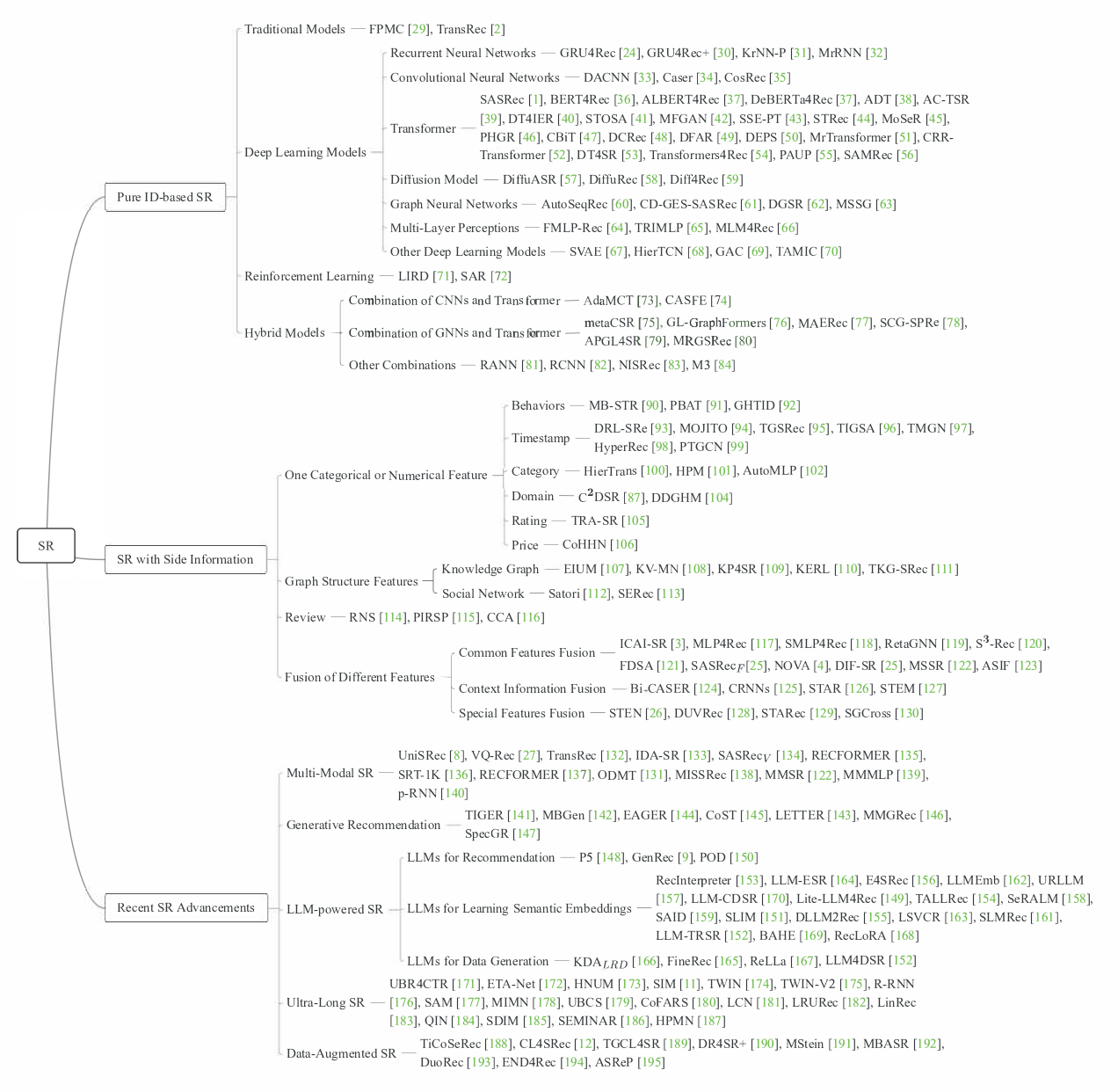}
		\caption{Taxonomy of existing works on sequential recommendation.}
		\label{fig:SR_taxonomy}
	\end{figure*}

	Our survey introduces SR models through a comprehensive taxonomy. As shown in Figure~\ref{fig:SR_taxonomy}, our survey will introduce SR models from three different aspects (i.e., pure ID-based SR, SR with side information and recent SR advancements). Our survey will discuss specific details about these SR models in following sections.
	\subsection{Challenges}
	In sequential recommender systems, there are some challenges as follows:
	\begin{itemize}
		\item Cold-start and data sparsity problems: A new user and a new item usually have sparse interactions. It is a challenge to improve recommendation performance in a new user and a new item.
		\item Ultra-long interaction sequences modeling: Over time, some users might generate ultra-long interaction sequences. It is a challenge to model these ultra-long interaction sequences effectively and efficiently.
		\item Dynamic interests: Users' interests evolve over time. It is a challenge to capture users' current and long-term interests.
		\item Diverse interests: Users exhibit diverse interests across different items. It is a challenge to capture users' diverse interests.
		\item Knowledge transfer across platforms: Items might be quiet different across different platforms. It is a challenge to achieve knowledge transfer across different platforms.
	\end{itemize}
	\section{Pure ID-based SR}\label{sec:pure_ID_based_SR}
	In SR models based on pure IDs, each item and user are represented by a unique item identity (i.e., item ID) and a unique user identity (i.e., user ID), respectively. In the past few years, the number of SR models based on pure IDs has increased significantly. We categorize them into different classes according to the used techniques. In this section, we will first introduce some SR methods using traditional machine learning. Then, we will introduce some SR methods integrating various deep learning techniques. Next, we will introduce some SR methods using reinforcement learning. Finally, we will introduce some SR methods that combine different kinds of deep learning techniques. In this section, all SR models are designed based on pure IDs.
	\subsection{Traditional Models}
	Traditional SR models do not use deep learning techniques. Though these models are simple and effective, they cannot capture long-term sequential dependencies. The traditional SR models can mainly be classified into three classes: frequent pattern mining, Markov models and latent factor models.
	
	FPM (frequent pattern mining) extracts different kinds of patterns from interaction sequences \cite{CSP}. These models are of good explainability, but they cannot tackle complex data. If a data is too large, the number of extracted sequential patterns might become excessively too large.
	
	Markov models have an assumption that a predicted action relies on a few recent actions. For example, FPMC \cite{FPMC} predicts the next interacted item for each user by considering both the last interacted item and personalized interests. However, these models cannot exploit long-term sequential dependencies. If the order dependencies are weak in data, Markov models might perform poorly.
	
	Latent factor models mainly learn latent representations to estimate unobserved transitions. TransRec \cite{TransRec_t} proposes a transition operator: a previously interacted item + a user $\approx$ the user next interacted item. However, the transition operator might vary across different data. Meanwhile, these SR models cannot capture complex transition relations.
	\subsection{Deep Learning Models}
	\subsubsection{Recurrent Neural Networks}
	Compared with traditional SR models, RNNs-based SR models not only can capture short-term sequential dependencies, but also can capture long-term sequential dependencies and non-linear dynamics. In recent years, some RNNs-based SR models have been proposed. GRU4Rec \cite{GRU4Rec} applies RNNs to capture long-term dependencies in session-based data. Based on GRU4Rec, GRU4Rec+ \cite{GRU4Rec+} proposes a novel negative item sampling and a novel ranking loss function. Some RNNs-based SR models are designed to capture each sequence independently, thus ignoring the global structure of all sequences. To solve this problem, KrNN-P \cite{KrNN-P} incorporates neighbor sequences into RNNs to capture global relationships while refining local ones. To consider the dependencies among users' sequences, MrRNN \cite{MrRNN} introduces manifold regularization into RNNs based on the multi-facets of collaborative filtering. However, these RNNs-based models have some shortcomings. For example, they cannot avoid vanishing gradients while handling ultra-long interaction sequences.
	\subsubsection{Convolutional Neural Networks}
	Convolutional neural networks (CNNs) have been widely used in image processing. They mainly consist of two operations (i.e., a convolution operation and a pooling operation). Compared with multi-layer perceptron (MLPs), CNNs can extract features from images more effectively. By regarding user $u$’s interaction sequence as an image, CNNs can also be applied to sequential recommendation. CNNs usually have multiple convolutional filters. Therefore, CNNs-based SR models can extract multi-faceted interests from user $u$'s interaction sequence.
	
	In recent years, several CNNs-based SR models have been proposed. DACNN \cite{DACNN} leverages CNNs to capture local sequential features while considering item sequences. Caser \cite{Caser} leverages convolutional filters to capture both general preferences and sequential patterns. CosRec \cite{CosRec} leverages 2D convolutional neural networks to model complex interaction relationships. However, CNNs-based SR models have certain shortcomings. For example, they might have a large number of convolutional filters, which will result in excessive computational resource consumption.
	\subsubsection{Transformer}
	Transformer has accomplished great achievements in natural language processing, largely due to its self-attention mechanism. The self-attention mechanism can identify crucial information by calculating attention scores.
	
	Transformer architecture has two types: single-direction Transformer (i.e., uni-Transformer) and double-direction Transformer (i.e., bi-Transformer). SASRec \cite{SASRec} is currently the most popular SR model, which leverages the uni-Transformer. SASRec predicts the next interacted items based on historical interacted items. BERT4Rec \cite{BERT4Rec} leverages bi-Transformer, which predicts the next interacted items according to historical interacted items and future interacted items. ALBERT4Rec and DeBERTa4Rec \cite{ALBERT4Rec_DeBERTa4Rec} demonstrate superior performance and training efficiency compared to BERT4Rec \cite{BERT4Rec}. ADT \cite{ADT} learns disentangled diverse interests using a Transformer-based encoder and decoder. However, the large weights of attention mechanism might not be accurate. Additionally, position encoding and noise inputs might negatively impact the recommendation performance. Therefore, AC-TSR \cite{AC-TSR} leverages contrastive learning to achieve noise reduction. DT4IER \cite{DT4IER} leverages decision Transformer to solve user retention challenges. Due to the uncertainty of users’ sequential behaviors, STOSA \cite{STOSA} leverages stochastic Gaussian distribution to embed items and introduces a novel Wasserstein self-attention module to capture item-item relationships. MFGAN \cite{MFGAN} uses a Transformer-based generator to recommend the next possible items and multiple discriminators to evaluate generated sub-sequences. SSE-PT \cite{SSE-PT} introduces user embeddings to realize the personalization of Transformer. STRec \cite{STRec} leverages sparse Transformer to focus on the most relevant interactions while predicting target items. Meanwhile, STRec \cite{STRec} replaces the self-attention mechanism with a cross-attention mechanism. MoSeR \cite{MoSeR} captures motifs hidden in behavior sequences to model micro-structure features. PHGR \cite{PHGR} leverages a novel hyperbolic inner product operator  to enable global (all user-item interactions) and local (each user’s interactions) graph representation learning in Poincaré ball. CBiT \cite{CBiT} leverages contrastive learning and bi-Transformer to learn more fine-grained preferences. DCRec \cite{DCRec} leverages cross-view contrastive learning to capture correlations among different users' interaction sequences. DFAR \cite{DFAR} leverages a dual-interest disentangling layer to disentangle positive and negative interests for learning users’ transition patterns better. DEPS \cite{DEPS} leverages both users’ and items' interaction sequences as the inputs to two Transformers for learning preferences better. MrTransformer \cite{MrTransformer} leverages a preference separation module to learn both users’ common and unique preferences. CRR-Transformer \cite{CRR-Transformer} applies a pre-trained transformer model to an online RL algorithm. Due to the uncertainty of interactions, DT4SR \cite{DT4SR} leverages an elliptical Gaussian distribution to describe items and adopts Wasserstein distance to measure the similarity among distributions. Transformers4Rec \cite{Transformers4rec} applies Transformer-based architectures for recommendation, such as GPT-2, XLNET, and Transformer-XL. PAUP \cite{PAUP} captures both short- and long-term patterns through a progressive attention distribution mechanism. In this model,  long behavior sequences are segmented into a series of sub-sequences using a down-sampling convolution module. SAMRec \cite{SAMRec} enhances the recommendation performance of Transformer from the perspective of loss geometry.
	
	Though Transformer-based models have achieved significant progress, these models still have some shortcomings. For example, Transformer-based SR models suffer from a quadratic computational complexity.
	\subsubsection{Diffusion Model}
	Diffusion model has been extensively applied to various generative tasks, such as image generation.
	
	In recent years, several SR models based on diffusion model have been proposed. DiffuASR \cite{DiffuASR} leverages a diffusion-based framework to generate high-quality interactions guaranteeing the similarity between generated and original sequences. DiffuRec \cite{DiffuRec} is the first work that applies the diffusion model to sequential recommendation. By corrupting and reconstructing target item embeddings, DiffuRec \cite{DiffuRec} injects uncertainty during the recommendation process. Diff4Rec \cite{Diff4Rec}  leverages the diffusion model to augment data by adopting a curriculum scheduling strategy, including interaction augmentation and objective augmentation. However, Diffusion-based SR models have some weaknesses. Compared with Transformer-based SR models, diffusion-based SR models require more computational resources and training time.
	\subsubsection{Graph Neural Networks}
	GNNs-based SR models can learn complex high-order interactions and capture collaborative information. Therefore, GNNs-based SR models have received widespread attention. AutoSeqRec \cite{AutoSeqRec} leverages an encoder and three decoders to reconstruct user-item interaction and item transition matrices. CD-GES-SASRec \cite{CD-GES-SASRec} leverages a causal graph to distinguish causal and non-causal transitions. DGSR \cite{DGSR} constructs a user-item interaction graph to capture personal preferences and an item-item transition graph to model transitional relationships between adjacent items. Meanwhile, DGSR \cite{DGSR} injects high-order connections into graphs. MSSG \cite{MSSG} transforms items within a sequence into a star graph. Meanwhile, MSSG introduces an additional internal node to capture global information. MSSG overcomes the over-smoothing issue and realizes a linear time complexity. However, GNNs-based SR models might not fully capture sequential patterns and could be computationally expensive to train.
	\subsubsection{Multi-Layer Perceptions}
	Compared with Transformer, leveraging MLPs to calculate is more efficient (i.e., a linear time complexity). In recent years, some MLPs-based SR models have been proposed. FMLP-Rec \cite{FMLP-Rec} incorporates learnable filters into MLPs architecture for reducing noise. TRIMLP \cite{TRIMLP} disables triangle neurons in MLPs to account for chronological order in users' interaction sequences. MLM4Rec \cite{MLM4Rec} learns users’ global and local preferences by incorporating convolution operators into MLPs. However, these MLPs-based SR models struggle to capture complex sequential patterns.
	\subsubsection{Other Deep Learning Models}
	Variational autoencoder (VAE) can learn compressed features from input embeddings. SVAE \cite{SVAE} combines both VAE and RNNs to capture sequential dependencies, with VAE extracting features at each time step.
	
	Temporal convolutional network (TCN) is a specialized type of convolutional neural network. Compared with RNNs, TCN is more computationally efficient. HierTCN \cite{HierTCN} combines both TCNs and RNNs to learn short-term interests and long-term interests in sequences.
	
	Capsule networks can recognize objects from different perspectives. GAC \cite{GAC} leverages capsule networks to model personalized item-level and factor-level sequential dependencies, and combines the two kinds of sequential dependencies to recommend items to users.
	TAMIC \cite{TAMIC} utilizes a time-aware dynamic routing algorithm and two kinds of time-aware voting gates to inject temporal information into sequential modeling.
	
	\begin{figure*}[h]
		\includegraphics[width=\linewidth]{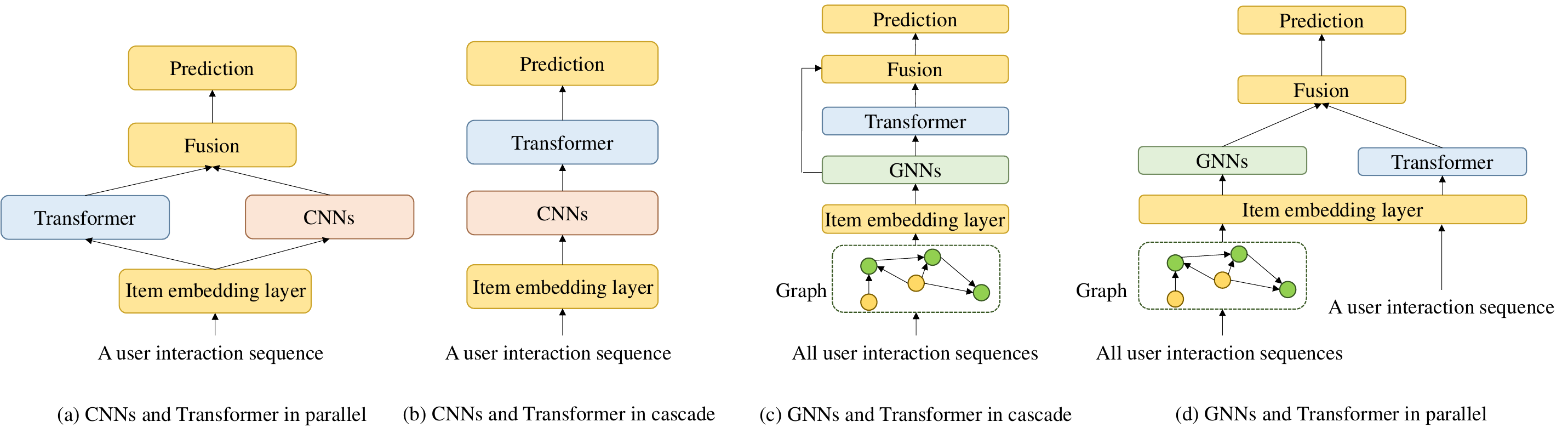}
		\caption{Illustration of the common combination strategies in sequential recommendation.}
		\label{fig:combined-type}
	\end{figure*}
	\subsection{Reinforcement Learning}
	Reinforcement learning can efficiently model interactions between users and a recommender system. LIRD \cite{LIRD} regards sequential interactions between the users and the recommender system as a Markov decision process (MDP), and leverages reinforcement learning to learn optimal strategies for recommending items automatically. SAR \cite{SAR} leverages reinforcement learning to learn dependencies and the length of the interaction sequence for each user.
	\subsection{Hybrid Models}
	Some SR models demonstrate significant improvements in recommendation performance by combining different techniques. For example, combining Transformer and GNNs can achieve better performance. As shown in Table~\ref{tab:different-methods}, each technique offers distinct advantages. Therefore, the hybrid models can combine the strengths of different techniques. The hybrid models can be roughly categorized into three classes (i.e., CNNs + Transformer, GNNs + Transformer and other combined types). Integrating various neural network architectures enables SR models to achieve enhanced recommendation performance. Four kinds of commonly used combined models are depicted in Figure~\ref{fig:combined-type}. Two different techniques can be leveraged in parallel. For example, CNNs and Transformer can extract short-term dependencies and long-term dependencies from user interaction sequences, respectively. And then the two kinds of dependencies are fused together. Alternatively, the two techniques can be used in a cascaded manner. For example, GNNs can be denoted as a feature extractor for extracting features, and then the extracted features can be fed into Transformer.
	\subsubsection{Combination of CNNs and Transformer}
	Transformer and CNNs exhibit unique abilities to capture different types of dependencies. Transformer has a good ability to capture long-term dependencies. CNNs are effective at modeling short-term dependencies and multiple-faced preferences. As shown in part (a) of Figure~\ref{fig:combined-type}, CNNs and Transformer can be combined in a parallel manner, which is formalized as follows:
	\begin{equation}
		\mathbf{P}_{u}^{CNNs} = \operatorname{CNNs}(\operatorname{Embed}(S_{u}))
	\end{equation}
	\begin{equation}
		\mathbf{P}_{u}^{Tr} = \operatorname{Transformer}(\operatorname{Embed}(S_{u}))
	\end{equation}
	\begin{equation}
		\mathbf{P}_{u}^{final} = \operatorname{Fuse}(\mathbf{P}_{u}^{CNNs},\mathbf{P}_{u}^{Tr})
	\end{equation}
	The final preferences $\mathbf{P}_{u}^{final}$ can be fed into a prediction module for recommending personalized items for user $u$. Meanwhile, as shown in part (b) of Figure~\ref{fig:combined-type}, CNNs and Transformer can be integrated in a cascaded manner, which is formalized as follows:
	\begin{equation}
		\mathbf{P}_{u}^{final} = \operatorname{Transformer}(\operatorname{CNNs}(\operatorname{Embed}(S_{u})))
	\end{equation}

	AdaMCT \cite{AdaMCT} captures long-term and short-term dependencies by incorporating locality inductive bias into the Transformer with a local convolutional filter. CASFE \cite{CASFE} first leverages CNNs to capture periodic information, and then leverages Transformer to evaluate the importance of different behaviors for target behaviors.
	\subsubsection{Combination of GNNs and Transformer}
	GNNs and Transformer excel at modeling graph structure and sequences, respectively. Transformer can capture temporal patterns and long-term dependencies. Meanwhile, GNNs can capture structural information and users' long-term preferences. Therefore, combining GNNs and Transformer enables SR models to achieve better performance. As shown in part (d) of Figure~\ref{fig:combined-type}, GNNs and Transformer can be combined in a parallel manner, which is formalized as follows:
	
	\begin{equation}
		\mathbf{P}_{u}^{GNNs} = \operatorname{GNNs}(\operatorname{Embed}(\mathcal{G}),u)
	\end{equation}
	\begin{equation}
		\mathbf{P}_{u}^{Tr} = \operatorname{Transformer}(\operatorname{Embed}(S_{u}))
	\end{equation}
	\begin{equation}
		\mathbf{P}_{u}^{final} = \operatorname{Fuse}(\mathbf{P}_{u}^{GNNs}, \mathbf{P}_{u}^{Tr})
	\end{equation}
	In the meantime, as shown in part (c) of Figure~\ref{fig:combined-type}, GNNs and Transformer can be integrated in a cascade manner, which is formalized as follows:
	\begin{equation}
		\mathbf{P}_{u}^{GNNs} = \operatorname{GNNs}(\operatorname{Embed}(\mathcal{G}),u)
	\end{equation}
	\begin{equation}
		\mathbf{P}_{u}^{Tr} = \operatorname{Transformer}(\operatorname{GNNs}(\operatorname{Embed}(\mathcal{G}),S_{u}))
	\end{equation}
	\begin{equation}
		\mathbf{P}_{u}^{final} = \operatorname{Fuse}(\mathbf{P}_{u}^{GNNs}, \mathbf{P}_{u}^{Tr})
	\end{equation}
	Compared with the former manner, it is noted that the latter leverages GNNs to extract item embeddings as the input of Transformer.

	metaCSR \cite{metaCSR} leverages Meta Learner to extract and propagate transferable knowledge from prior users and effectively learn presentations for cold-start users. GL-GraphFormers \cite{GL-GraphFormers} leverages a global user-item bipartite graph to learn both first- and second-order graph information, and then injects the information into Transformer in the form of input and attention weights. MAERec \cite{MAERec} adaptively and dynamically distills sequential transition patterns for constructing a contrastive learning framework. SCG-SPRe \cite{SCR-SPRe} leverages a complement graph and a substitute graph to encode the complementary and substitutable relations, respectively. To explore global item-to-item interactions (collaborative information), APGL4SR \cite{APGL4SR} constructs a global item graph. Meanwhile, to capture personalized patterns, user embeddings are leveraged. MRGSRec \cite{MRGSRec} leverages a sequential encoder and a graph encoder to learn local behavioral representations and global behavioral representations, respectively. Subsequently, the two preferences are fused together for final recommendation. However, these methods have some drawbacks. For example, learning collaborative signals through graph neural networks is computationally expensive.
	\subsubsection{Other Combinations}
	There are also some SR models combining various other techniques. RANN \cite{RANN} leverages the strengths of RNNs and the self-attention mechanism in Transformer to capture user preferences from different views. RCNN \cite{RCNN} combines the advantages of RNNs and CNNs, in which RNNs can capture long-term dependencies and CNNs can capture short-term sequential patterns. NISRec \cite{NISRec} captures and aggregates the long-term and short-term intentions of neighbor users. Then NISRec \cite{NISRec} fuses these intentions for recommendation. Mixture networks can leverage multiple experts to learn diverse interests. For example, M3 \cite{M3} leverages a mixture of experts, where each expert specializes in capturing a specific temporal range. These experts are dynamically combined through a learned gating mechanism. Therefore, M3 \cite{M3} can capture diverse behavior patterns effectively.
	
	By combining different techniques, SR models can achieve better performance. However, these models have several drawbacks. Firstly, they are more complex and computationally expensive. Secondly, different techniques might capture conflicting preferences. For example, Transformer and GNNs capture distinct preferences, which may conflict. Thirdly, different techniques might converge at different rates.
	\section{SR with Side Information}\label{sec:SR_with_side_information}
	In this section, our survey first discusses SR models combining IDs and one of the general features (i.e., categorical features and numerical features). Then, our survey shows SR models combining IDs and one of the graph-structure features (i.e.,  knowledge graph and social networks). Next, our survey introduces SR models combining IDs and reviews. Finally, our survey shows SR models combining IDs and multiple feature types.
	\begin{table*}[h]
		\caption{The advantages and disadvantages of different SR models.}
		\label{tab:different-methods}
		\centering
		\begin{tabular}{p{4cm}p{6cm}p{6cm}}
			\hline
			SR models & Advantages  & Disadvantages\\
			\hline
			Traditional SR models  & Simple, can capture short-term sequential dependencies, efficient, good explainability & cannot capture long-term sequential dependencies, cannot handle complex data \\
			RNNs-based SR models & Can capture long-term sequential dependencies & Are sensitive to the order of items in an interaction sequence\\
			CNNs-based SR models & Can model union-level sequential patterns, are not sensitive to the order of interacted items & High computational costs due to lots of convolutional filters\\
			Transformer-based SR models & Can identify the importance of interacted items, can capture short-term and long-term dependencies & Time complexity is related to the length of interaction sequences\\
			Diffusion-based SR models & Can alleviate data sparsity, can generate diversity interaction data & Complex training process, high computational costs\\
			GNNs-based SR models & Can model complex transitions among interacted items, can capture structure characteristics and collaborative signals from a global view & High computational costs, over-smoothing problem\\
			MLPs-based SR models & Computational efficiency, a linear time and space complexity, simple model architecture &  Limited performance on ultra-long sequences, are weak in capturing sequential dependencies \\
			VAE-based SR models & Can capture complex latent relationships, can handle the uncertainty of users' preferences & High computational costs, are influenced easily by potential noise\\
			TCNs-based SR models & Can preserve sequence order during convolution process and conduct efficient sequence modeling & Are difficult to capture long-term sequential dependencies, high complexity for hyper-parameter searching\\
			Capsule-networks-based SR models & Can model multi-facet interests, can model complex interaction sequences & Are difficult to capture long-term sequential dependencies\\
			RL-based SR models & Can adapt to changing sequence lengths, can model interaction process & cannot solve the cold-start problem, rely heavily on real-time feedback\\
			\hline
		\end{tabular}
	\end{table*}
	\subsection{One Categorical or Numerical Feature}
\begin{figure*}[h]
		\includegraphics[width=\linewidth]{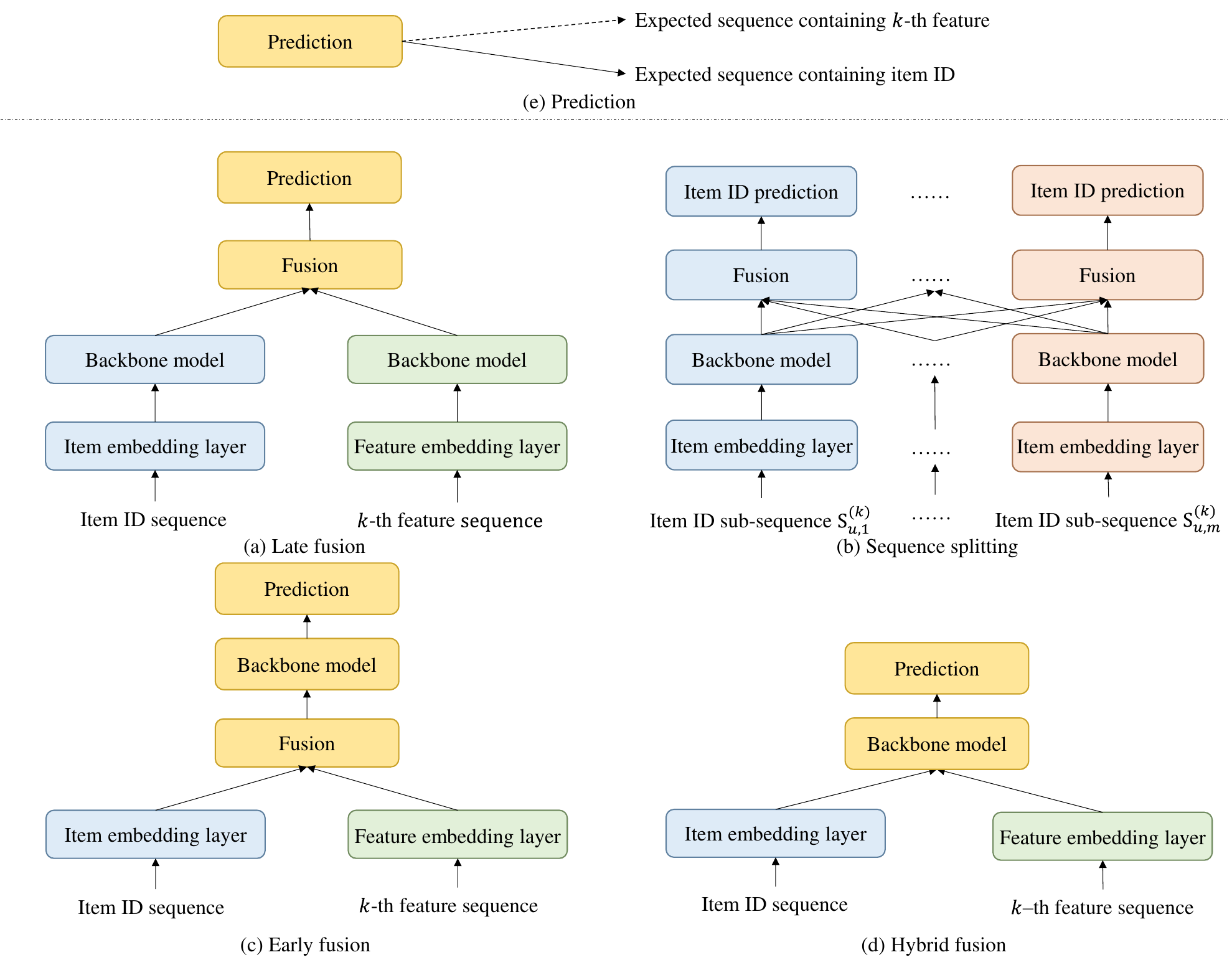}
		\caption{Illustration of the usage of one categorical or one numerical feature in sequential recommendation.}
		\label{fig:one-sparse-feature}
	\end{figure*}

	Many SR models combine either a categorical or a numerical feature with item IDs to recommend items for users. Some specific methods are shown in Figure~\ref{fig:one-sparse-feature}. From Figure~\ref{fig:one-sparse-feature}, we can know that these methods can be grouped into four main types. As shown in part (a) of Figure~\ref{fig:one-sparse-feature}, the first method leverages two backbones to learn two kinds of preferences from one sequence containing item IDs and the other sequence containing feature $k$, respectively. These preferences are then fused using methods such as contrastive learning, concatenation, addition, gate addition, or a cross-attention mechanism. The fusion process is formalized as follows:
	
	\begin{equation}
		\mathbf{P}_{u}^{final} = \operatorname{Fuse}(\operatorname{BM}_1(\mathbf{E}_{u}), \operatorname{BM}_2(\mathbf{A}_{u}^{(k)}))
	\end{equation}
	As shown in part (b) of Figure~\ref{fig:one-sparse-feature}, the second method leverages the value of feature $k$ to split the original sequence containing item IDs into sub-sequences. For example, some SR models split hybrid interaction sequences into separate interaction sequences in each domain \cite{CD-SASRec,Tri-CDR}. Different preferences are then extracted from these sub-sequences and fused together to predict personalized items. The fusion process is formalized below:
	\begin{equation}
		\begin{aligned}
		\mathbf{P}_{u}^{final} =& \operatorname{Fuse}(\operatorname{BM}_{1}(\operatorname{Embed}(S_{u,1}^{(k)})),\ldots, \\
		&\operatorname{BM}_{m}(\operatorname{Embed}(S_{u,m}^{(k)})))
		\end{aligned}
	\end{equation}
	It is noted that the fusion usually exhibits a certain directionality, where user preferences are transferred from a source domain to a target domain \cite{C2DSR}.
	As shown in part (c) of Figure~\ref{fig:one-sparse-feature}, the third method first leverages simple techniques to combine item embeddings and feature embeddings. Then, the fused embeddings are fed to a backbone model for recommendation. For example, adding them together is to obtain the input of a backbone model. The fusion process is formalized as follows:
	\begin{equation}
		\mathbf{P}_{u}^{final} = \operatorname{BM}_1(\operatorname{Fuse}(\mathbf{E}_{u}, \mathbf{A}_{u}^{(k)}))
	\end{equation}
	As shown in part (d) of Figure~\ref{fig:one-sparse-feature}, the last method integrates item embeddings with feature embeddings in a backbone model. For example, leveraging feature embeddings changes the weights in attention matrices. The fusion process is formalized as follows:
	
	\begin{equation}
		\mathbf{P}_{u}^{final} = \operatorname{BM}_1(\mathbf{E}_{u}, \mathbf{A}_{u}^{(k)})
	\end{equation}
	Though all four kinds of methods can obtain final user $u$'s preferences, they vary in terms of complexity and efficiency \cite{C2DSR,TiSASRec}. Meanwhile, as shown in part (e) of Figure~\ref{fig:one-sparse-feature}, the predicted goal might contain the feature of the next interacted item \cite{TimelyRec,DIF-SR}.
	\subsubsection{Behaviors}
	Different behaviors show users' different preferences. Click behaviors usually reflect users' short-term preferences. Favorite behaviors usually reflect users' middle-term preferences. Purchase behaviors usually reflect users' long-term preferences. Meanwhile, dependencies exist among these different behaviors. MB-STR \cite{MB-STR} injects users’ behaviors into Transformer to capture multi-behavior heterogeneous dependencies. PBAT \cite{PBAT} models users’ personalized behavior patterns and behavioral collaborations by a personalized behavior pattern generator and a behavior-aware collaboration extractor, respectively. GHTID \cite{GHTID} leverages a global heterogeneous graph (constructed from all interaction sequences) and some local heterogeneous graphs (constructed from each interaction sequence) to explicitly learn heterogeneous item transitions.
	\subsubsection{Timestamp}
	Users usually exhibit certain behaviors at specific times. For example, users tend to purchase T-shirts in summer. By considering timestamps, SR models can capture users' evolving interests more precisely. DRL-SRe \cite{DRL-SRe} exploits time-sliced graph neural networks to learn complex user-item interactions from a global perspective, and leverages GRUs to learn complex item-to-item correlations. To account for temporal context in recommendations, MOJITO \cite{MOJITO} leverages a Gaussian mixture method to combine attention-based temporal context with item representations. TGSRec \cite{TGSRec} leverages a temporal collaborative Transformer layer to capture collaborative signals and temporal dynamics. TIGSA \cite{TIGSA} leverages a time interval-aware graph to consider the impact of time intervals for recommendation, and leverages Transformer to extract users' sequential preferences. TMGN \cite{TMGN} first identifies the $k$ most similar users to each candidate user. And then, TMGN \cite{TMGN} injects multiple users' sequences and temporal information into multi-head attention. By leveraging timestamps, HyperRec \cite{HyperRec} splits a whole user-item interaction hypergraph into sub-hypergraphs. HyperRec \cite{HyperRec} adopts these sub-hypergraphs and multiple convolutional layers to capture multi-order connections and short-term user intents. Then the dynamic item embeddings and short-term user intents are incorporated by a fusion layer, and the incorporated representations are fed into a self-attention layer. PTGCN \cite{PTGCN} leverages a position-enhanced and time-aware graph convolution operation to capture sequential patterns and temporal dynamics.
	\subsubsection{Category}
	The category of interacted items can reflect users' preferences more precisely. HierTrans \cite{HierTrans} leverages a novel hierarchical temporal graph to extend traditional item-level relations to category-level relations. HPM \cite{HPM} leverages Transformer to learn both low-level preferences (i.e., item IDs) and high-level preferences (i.e., item categories). While considering both item IDs and the categories of items, AutoMLP \cite{AutoMLP} captures users’ long-term and short-term interests, with short-term interests learned by an adaptive search algorithm.
	\subsubsection{Domain}
	Domain information plays a crucial role in sequential recommender systems. SR models usually first handle separate interaction sequences in each domain. Then SR models transfer the obtained knowledge from a source domain to a target domain \cite{Tri-CDR, MGCL}. It is because data distribution is different across domains. C$^{2}$DSR \cite{C2DSR} constructs a global item-to-item graph across two domains to learn transitions between two domains. DDGHM \cite{DDGHM} constructs local graphs and a global graph to capture intra-domain sequential transitions and inter-domain sequential transitions, respectively. And, DDGHM leverages hybrid metric learning to alleviate data sparsity and improve recommendation performance.
	\subsubsection{Rating}
	Users usually rate interacted items, indicating their preferences. By considering ratings, users' preferences can be more accurately captured. TRA-SR \cite{TRA-SR} integrates rating information into the weight calculation of self-attention to enhance the learning of users' preferences.
	\subsubsection{Price}
	In real life, users often prefer purchasing cheaper items. Even if users have a strong liking for an item, they may not purchase it if it is too expensive. Both price and interest preferences might influence users' purchase choices simultaneously. Therefore, CoHHN \cite{CoHHN} designs a co-guided heterogeneous hypergraph network to simultaneously extract users' price preferences and interest preferences. Then, CoHHN leverages a co-guided learning schema to model the complex relations between the two kinds of preferences. Finally, CoHHN predicts users' behaviors based on the two kinds of preferences and item features.
	\subsection{Graph Structure Features}
	\subsubsection{Knowledge Graph}
	Knowledge graph is a structure representation of knowledge, where entities (nodes) are connected by relationships (edges). The entities can represent users, items and other related features. The relationships represent meaningful connections (e.g., “a user purchases a book" and “The color of the iPhone is black"). Therefore, knowledge graph can describe the relations among items, users, and features, as well as interactions between users and items. Knowledge graph provides rich extra knowledge for recommending personalized items. Therefore, by leveraging knowledge graph, SR models can alleviate cold-start and data sparsity problems. Meanwhile, SR models based on knowledge graph are also capable of providing explanations for recommendation.
	
	There are several SR models based on knowledge graph. EIUM \cite{EIUM} leverages knowledge graph to conduct explainable recommendation. KV-MN \cite{KV-MN} integrates knowledge graph into RNNs to enhance semantic representations. KP4SR \cite{KP4SR} leverages knowledge graph to capture users’ fine-grained preferences. Meanwhile, KP4SR \cite{KP4SR} converts knowledge graph to knowledge instructions for noise reduction. KERL \cite{KERL} incorporates knowledge graph into reinforcement learning. In order to learn sequential dependencies while considering semantic information, KERL \cite{KERL} leverages a composite reward function to compute both sequence-level and knowledge-level rewards. TKG-SRec \cite{TKG-SRec} leverages temporal knowledge graph to capture temporal information, and leverages GRUs to conduct temporal knowledge evolution training. However, SR models based on knowledge graph have several disadvantages as follows: 1) Complexity: Knowledge graph contains many entities and relationships, which may increase the complexity of SR models and cost more computational resources. 2) Noise: Constructing a high-quality knowledge graph is challenging. Incorrect relationships within the knowledge graph might degrade the recommendation performance.
	\subsubsection{Social Network}
	Social networks provide insight into the friend relationships among users. Since users tend to have similar preferences to their friends, leveraging social networks allows SR models to capture users' preferences more accurately. Satori \cite{Satori} leverages a graph attention network and a self-attention mechanism to extract auxiliary features and user intentions, respectively. Then, hybrid user and item representations are sent to a prediction module to calculate predicted scores. SERec \cite{SERec} first constructs a heterogeneous graph by combining a social network and users' interaction sequences. Then by leveraging GNNs, SERec improves user and item representations by integrating knowledge from the social network.
	\subsection{Review}
	Compared with rating information, reviews can reflect users' preferences about items comprehensively in the form of text and images. Meanwhile, reviews can also provide more detailed information about items. By reading reviews given by other people, users' action decisions might be influenced. RNS \cite{RNS} learns users' long-term preferences by an aspect-aware convolutional network over users' document. Meanwhile, RNS learns users' short-term preferences by hierarchical attention. Finally, RNS combines the two kinds of preferences for recommendation. To consider both user-item interactions and reviews simultaneously, PIRSP \cite{PIRSP} learns item sequential patterns and review sequential patterns together. Then PIRSP introduces a fusion gating mechanism to learn short-term preferences from the two patterns. Finally, PIRSP combines short-term and long-term preferences for recommendation. CCA \cite{CCA} designs a cascaded cross-attention mechanism to capture the complex relationships among an item sequence, a review sequence and candidate items.
	\subsection{Fusion of Different Features}
	\begin{figure}[h]
		\includegraphics[width=\linewidth]{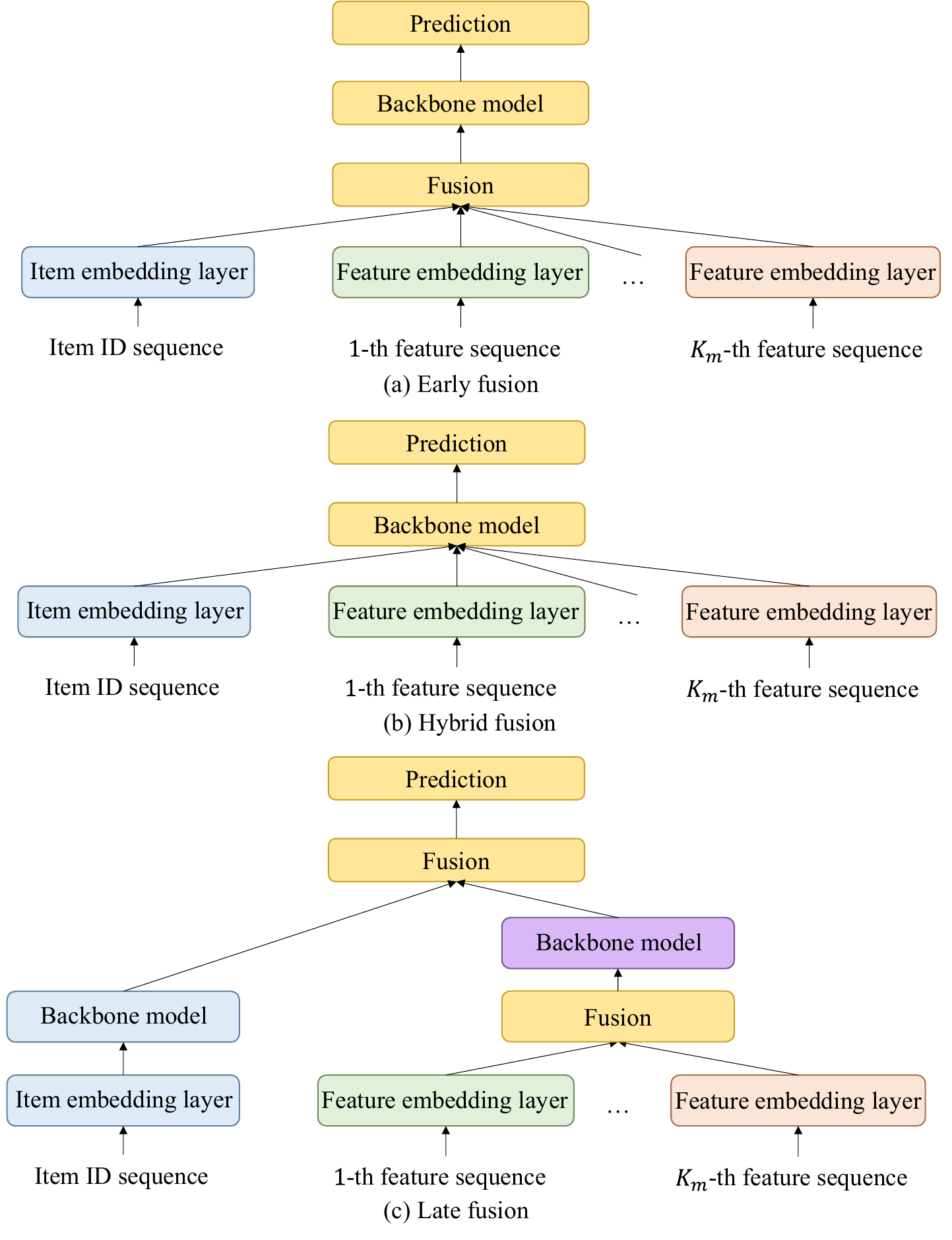}
		\caption{Illustration of different fusion strategies for multiple general features.}
		\label{fig:multi-features-fusion}
	\end{figure}
	In SR models, multiple features are usually used to improve the recommendation performance. Different features can be fused by different fusion methods. These fusion methods can mainly be summarized into three types. Some specific details are shown in Figure~\ref{fig:multi-features-fusion}. The first type leverages simple methods to fuse item embeddings and feature embeddings together, for example, adding them together is to obtain the input of a backbone model. The fusion is shown as follows:
	\begin{equation}
		\mathbf{P}_{u}^{final} = \operatorname{BM}_1(\operatorname{Fuse}(\mathbf{E}_{u},\mathbf{A}_{u}^{(1)},\ldots,\mathbf{A}_{u}^{(K_{g})}))
	\end{equation}

	As shown in part (b) of Figure \ref{fig:multi-features-fusion}, the second type fuses different features in a backbone model \cite{DIF-SR}, for example, leveraging a decoupled features fusion mechanism in a self-attention mechanism. The fusion is shown as follows:
		\begin{equation}
			\mathbf{P}_{u}^{final} = \operatorname{BM}_1(\mathbf{E}_{u},\mathbf{A}_{u}^{(1)},\ldots,\mathbf{A}_{u}^{(K_{g})})
	\end{equation}

	As shown in part (c) of Figure \ref{fig:multi-features-fusion}, the third type first fuses different feature embeddings together, and then passes the fused embeddings through a backbone model to learn feature representations. Meanwhile, item representations are learned by the other backbone model. Finally, the item representations and feature representations are fused together for prediction. The fusion is described as follows:
	\begin{equation}
		\mathbf{P}_{u}^{feat} = \operatorname{BM}_1(\operatorname{Fuse}(\mathbf{A}_{u}^{(1)},\ldots,\mathbf{A}_{u}^{(K_{g})}))
	\end{equation}
	\begin{equation}
		\mathbf{P}_{u}^{final} = \operatorname{Fuse}(\operatorname{BM}_2(\mathbf{E}_{u}),\mathbf{P}_{u}^{feat})
	\end{equation}

	We will introduce common features fusion, context information fusion and special features fusion one by one.
	\subsubsection{Common Features Fusion}
	In this section, we mainly introduce some SR models that fuse some common features (i.e., position, category, brand, etc.) and item IDs. The structure of these SR models is feature-agnostic to some extent. ICAI-SR \cite{ICAI-SR} leverages a heterogeneous graph to capture complex item-to-attribute relationships by  inner attribute aggregation and attribute-to-item aggregation, and then passes aggregation embeddings through an entity sequential model. MLP4Rec \cite{MLP4Rec}  designs a fusion mechanism to capture sequential, cross-channel and cross-feature correlations. SMLP4Rec \cite{SMLP4Rec}  leverages layer normalization for a sequence mixing module, a feature mixing module and a channel mixing module simultaneously. Meanwhile, to enhance efficiency and effectiveness, SMLP4Rec leverages a parallel mode. RetaGNN \cite{RetaGNN} leverages a relational attentive GNN in which learnable weight matrices focus on various relations among users, items and attributes to have inductive and transferable properties. Then, RetaGNN \cite{RetaGNN} leverages Transformer to capture short-term and long-term temporal patterns. S$^{3}$-Rec \cite{S3-Rec} leverages four self-supervised learning tasks (i.e., associated attribute prediction, masked item prediction, masked attribute prediction and segment prediction) to learn the associations between item IDs and item features. 
	
	In Transformer-based SR models, several advancements have been made in combining common features with item IDs. FDSA \cite{FDSA} first fuses different features by a vanilla attention mechanism. Then, FDSA leverages separate self-attention blocks to independently learn feature-level and item-level representations. Finally, FDSA fuses the two kinds of preferences for recommendation. SASRec$_{F}$ \cite{DIF-SR} is a sample model to fuse common feature embeddings and item ID embeddings through a concatenation operation. This fusion method is considered invasive \cite{NOVA}, similar to an addition operation and a gating addition operation. The specific process is shown in part (a) of Figure \ref{fig:multi-features-fusion}. This fusion method might compromise the original information in item embeddings, potentially leading to negative effects. To address this issue, NOVA \cite{NOVA} leverages a non-invasive method to enhance the learning of attention matrices. In NOVA, key and query matrices are derived by integrating different feature embeddings and item embeddings while value matrices are directly obtained from item embeddings. However, NOVA \cite{NOVA} still leverages integrated embeddings to learn key and value matrices. The compounded embedding space might introduce random disturbances and share identical gradients, which might degrade recommendation performance. DIF-SR \cite{DIF-SR} leverages decoupled attention to adaptively learn different features by flexible gradients. MSSR \cite{MMSR} captures the associations among an item ID sequence and each feature sequence. ASIF \cite{ASIF} maintains semantic consistency between item IDs and features. Meanwhile, ASIF \cite{ASIF} makes full use of different kinds of features without compromising item embeddings, thanks to contrastive learning.
	\subsubsection{Context Information Fusion}
	Context information can reflect the specific environment where behaviors happen. SR models can provide more accurate recommendations by considering the context information. Bi-CASER \cite{Bi-CASER} uses a context graph to represent context information, and then leverages bi-Transformer to learn users’ sequential behaviors. CRNNs \cite{CRNNs} incorporates context information (e.g., interaction timestamps) by combining item embeddings with context embeddings to recommend personalized items. STAR \cite{STAR} leverages stacked RNNs to model the sequential and context-aware information simultaneously. STEM \cite{STEM} considers item behaviors and leverages a Transformer-based architecture for exploiting spatial and temporal information.
	\begin{figure*}[h]
		\includegraphics[width=\linewidth]{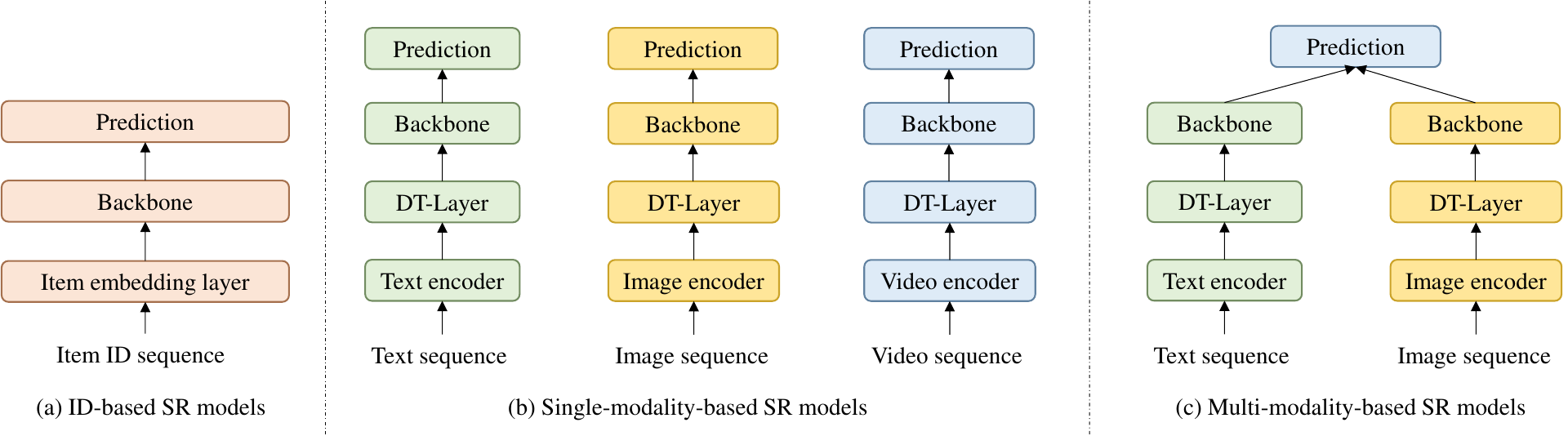}
		\caption{Illustration of different modal-based SR models. Note that DT denotes the dimension Transformer.}
		\label{fig:modal-based-SR-methods}
	\end{figure*}
	\subsubsection{Special Features Fusion}
	In this section, we mainly introduce SR models designed to fuse special features. These SR models are typically applied to situations where certain special features (e.g., social networks and timestamps) are available. STEN \cite{STEN} introduces temporal information and a social network to capture users’ fine-grained dynamic interests in an event-level direct paradigm. DUVRec \cite{DUVRec} constructs a timespan-aware sequence graph and an attribute-augmented graph to learn item-view and factor-view user representations, respectively. Recommendation performance is then improved by fusing the two representations. Based on features and labels, STARec \cite{STARec} retrieves a target user’s historical behaviors through a search-based retriever. Then, STARec \cite{STARec} leverages a graph structure to enhance representation learning. SGCross \cite{SGCross}  transfers knowledge from an auxiliary domain to a target domain from a personal view, a temporal view and a collaborative view. The knowledge encompasses personal preferences, dynamic preferences and collaborative preferences.
	\section{Recent SR Advancements}\label{sec:recent_SR_advancements}
	In this section, we will introduce recent research directions in sequential recommendation (SR) models. Firstly, we will discuss SR models using multi-modal features. Secondly, we will explore a novel paradigm (i.e., generative recommendation) in sequential recommendation. Thirdly, we will introduce SR models using LLMs. Fourthly, we will introduce SR models designed to address ultra-long interaction sequences. Finally, we will present augmentation techniques aimed at enhancing SR performance.
	\subsection{Multi-Modal SR}\label{sec:multi_modal_SR}
	A user $u$ can be represented by the user’s profile. The profile may consist of user $u$’s demographic information or user $u$’s interacted items. An item can be denoted by an item ID embedding or a pre-trained modality embedding. Modality content features consist of text, images, videos, audio, text-image multi-modal pairs, and so on. ID-agnostic representations can be learned using pre-trained modality models. A common text encoder is BERT, which is usually used to extract semantic features from text. Swin Transformer is an image encoder, which can be leveraged to extract features from images. As shown in Figure~\ref{fig:modal-based-SR-methods}. unlike ID-based SR models, one type of modal-based SR models leverages a single modality feature (e.g., text or images) to denote items. The other type of modal-based SR models leverages some modal features (e.g., text and images) to represent items \cite{ODMT}. In order words, the representations of item $i$ are obtained from the multi-modal features of item $i$ in modal-based SR models, which is formalized as follows:
	\begin{equation}
		\mathbf{e}_i=\left\{\begin{array}{l}
			\operatorname{TextEncoder}(\text{Text}_{i}) \\
			\operatorname{ImageEncoder}(\text{Image}_{i}) \\
			\operatorname{VideoEncoder}(\text{Video}_{i})
		\end{array}\right.
	\end{equation}
	
	SR problems can be solved by NLP models. Different from natural language, interacted items do not typically follow a specific order in user interactions. If we swap two items in a user's interaction sequence, the user’s preferences might not be changed. If we swap two words in a sentence, the meaning of this sentence will be changed. Text contains rich semantic information in the real world. By using natural language like text, semantic gaps across different domains and platforms can be bridged. Therefore, SR models can be trained in a universal semantic space.
	
	There are some single-modality-based SR models. UniSRec \cite{UniSRec} leverages an MoE-enhanced adaptor and parametric whitening for domain adaptation. To learn universal sequence representations better, UniSRec proposes a seq-item contrastive task and a seq-to-seq contrastive task. The model is pre-trained in one domain and subsequently fine-tuned in another domain using either an inductive or a transductive approach. Meanwhile, UniSRec adopts two augmentation strategies (i.e., item drop and word drop). VQ-Rec \cite{VQ-Rec} is a novel approach to learn vector-quantized item representations for solving the problem about over-emphasizing the effect of text features, which might lead to poor recommendation performance. VQ-Rec converts item text to a series of codes. Then with a lookup over a code embedding table, VQ-Rec can obtain item embeddings. Based on representation schema, VQ-Rec leverages contrastive learning among the next ground-truth items and negative items (i.e., semi-synthetic negative items and mixed-domain negative items). Finally, based on the code-embedding method, VQ-Rec implements a novel cross-domain fine-tuning method. TransRec \cite{TransRec_n} is pre-trained in one domain, and then is fine-tuned in another domain. TransRec \cite{TransRec_n} achieves excellent recommendation performance. It is because that TransRec \cite{TransRec_n} first obtains knowledge from a source domain, and then transfers the obtained knowledge to a target domain. The three models mentioned above share a similar approach: they are initially pre-trained in one domain and then fine-tuned in another domain. IDA-SR \cite{IDA-SR} adopts a pre-trained text encoder to learn item representations by encoding item text descriptions. The pre-training tasks consist of next-item prediction, masked item prediction and permuted item prediction. These item representations can then be used for personalized recommendations through semantic transfer. SASRec$_{V}$ \cite{SASRecV} leverages videos to denote an item. Compared with SASRec using an item ID to denote an item, SASRec$_{V}$ \cite{SASRecV} can achieve comparable performance using an end-to-end framework. RECFORMER \cite{RECFORMER2} mainly focuses on embedding initialization. By adopting behavior-tuned pre-trained language models for the initialization of ID-based SR models, RECFORMER \cite{RECFORMER2} can achieve better recommendation performance without introducing additional computational costs. SRT-1K \cite{SRT-1K} calculates item embeddings using a trainable feature extractor, which can make the number of parameters be independent of the number of items. Additionally, SRT-1K \cite{SRT-1K} leverages contrastive learning to enhance catalog diversity representation. RECFORMER \cite{RECFORMER} flattens item attributes into sentences to represent items. The SR models mentioned above achieve improved recommendation performance by combining item IDs and multi-modal features.
	
	There are some multi-modality-based SR models. ODMT \cite{ODMT} fuses different modal information (i.e., text and images) together. Meanwhile, ODMT uses knowledge distillation to model weights for online recommendations. MISSRec\cite{MISSRec} leverages a multi-modal pre-training and transfer learning framework to integrate multi-modal information into SR models. MMSR \cite{MMSR} integrates multi-modal information into nodes. MMMLP \cite{MMMLP}  incorporates multi-modal data into MLPs to enhance recommendation performance. p-RNN \cite{p-RNN} leverages four different frameworks to combine some item features (i.e., images and text) and item IDs for modeling interaction sequences. Based on the aforementioned modal-based SR models, the following conclusions can be drawn:
	\begin{itemize}
		\item Modal-based SR models with an end-to-end training framework achieve comparable recommendation performance with pure ID-based SR models. Because the pre-trained features might not be suitable for recommendation.
		\item Modal-based SR models with larger parameters tend to have better recommendation performance than the counterpart containing smaller parameters in recommendation tasks.
		\item In a cold-start setting, modal-based SR models can obtain better recommendation performance than pure ID-based SR models.
		\item If we leverage a pre-trained modality model to extract features, we can obtain better performance than training a model from scratch.
	\end{itemize}
	In the meantime, the modal-based SR models have some advantages as follows:
	\begin{itemize}
		\item Leveraging modality information to denote an item has the advantages of interpretability and visualization.
		\item Modal-based SR models can alleviate the cold-start problem to some extent.
		\item Modal-based SR models can realize the transferability of the models across different platforms.
		\item Multi-modal features act as a bridge to better integrate recommendation and search algorithms.
	\end{itemize}
	The modal-based SR models also have several disadvantages as follows:
	\begin{itemize}
		\item Modal-based SR models require substantial computational resources and extended training time.
		\item Hyper-parameters need to be searched carefully, as improper hyper-parameters easily lead to model collapse.
		\item Semantic information from different modalities might not be suited for recommendation.
	\end{itemize}
	
	\subsection{Generative Recommendation}
	\begin{figure}[h]
		\includegraphics[width=\linewidth]{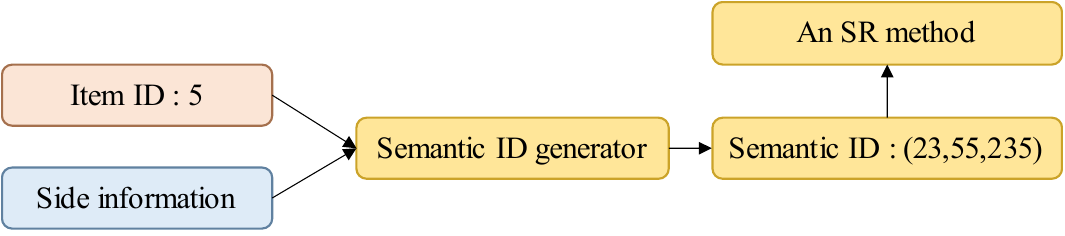}
		\caption{Illustration of semantic ID generation.}
		\label{fig:semantic-id}
	\end{figure}
	\begin{figure}[h]
		\includegraphics[width=\linewidth]{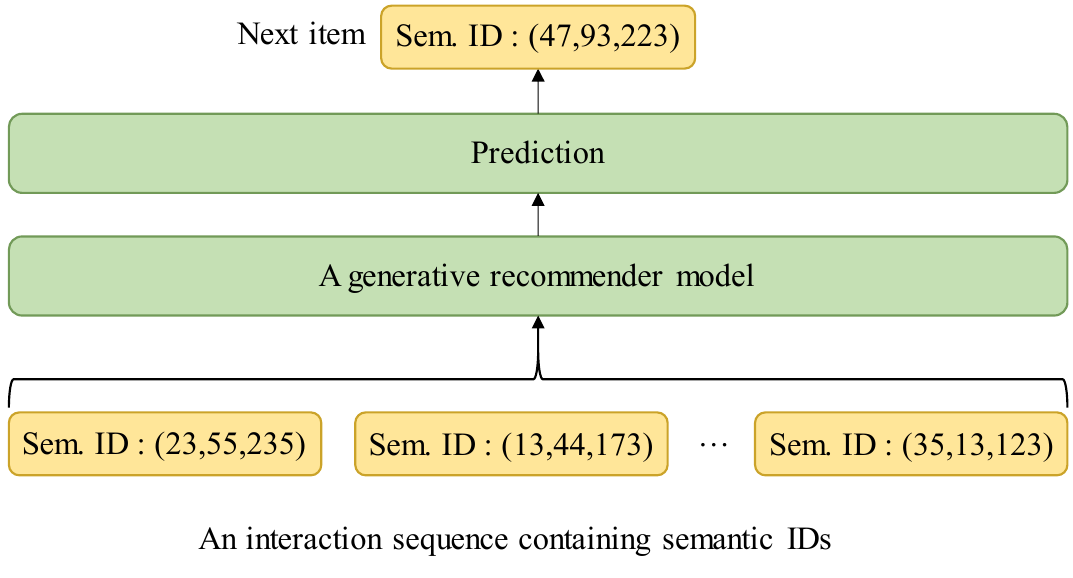}
		\caption{Illustration of generative recommendation with semantic IDs.}
		\label{fig:generative-recommend}
	\end{figure}
	Generative recommendation is proposed on the basis of a widely used technique in document retrieval (i.e., generative retrieval) \cite{TIGER}. Different from some above-mentioned SR models,	generative recommendation models \cite{MBGen, TIGER} leverage semantic IDs to denote an item. As depicted in Figure~\ref{fig:semantic-id}, a semantic ID consists of a series of tokens. Compared with only using an item ID to denote an item, this method has several advantages as follows:
	\begin{itemize}
		\item Compared with item IDs, semantic IDs encode semantic information extracted from side information (e.g., text descriptions) more efficiently.
		\item Because the semantic IDs of different items might contain some of the same tokens, knowledge can be shared among similar items.
		\item Using semantic IDs prevents the embedding table size from increasing linearly with the number of items in a data. It is because an item is denoted by a series of tokens instead of a unique item ID. The total number of tokens is far smaller than the number of all item IDs.
	\end{itemize}
	Some above-mentioned methods leverage the multi-modal features of an item to denote the item. However, these methods might damage fine-grained semantic information \cite{LETTER}. Semantic IDs can solve this problem. Due to their hierarchical nature, semantic IDs contain semantic information that ranges from a coarse-grained level to a fine-grained level. Generative recommendation generally involves: semantic ID generation and generative recommendation with semantic IDs. As depicted in Figure~\ref{fig:semantic-id}, semantic IDs are obtained by encoding item IDs and side information about items. RQ-VAE is a widely-used semantic ID generator, which leverages a residual quantizer to learn quantized representations \cite{TIGER}. Because target items are denoted by semantic IDs, a generative recommender model need to generate the target items by an auto-regressive manner. A widely-used generative recommender model is based on a Transformer-based encoder-decoder architecture \cite{EAGER,CoST}. The process is shown in Figure ~\ref{fig:generative-recommend}. In generative recommender models, the probability of user $u$ interacts with item $i$ is calculated by chain rule, which is formalized as follows:
	\begin{equation}
		P(i|S_{u})=\prod_{j=1}^l p\left(y_{i,j}|S_{u},y_{i,1},y_{i,2},\ldots,y_{i,j-1}\right)
	\end{equation}
	
	In recent years, some generative sequential recommender models have been proposed. TIGER \cite{TIGER} is the first model that leverages semantic IDs for generative recommendation. MBGen \cite{MBGen} considers the role of behavioral information in generative recommendation. To consider the complementary nature between behavioral information and semantic information, EAGER \cite{EAGER} leverages a shared encoder and two separate decoders to decode the two kinds of information. CoST \cite{CoST} leverages a contrastive quantization-based semantic tokenization approach to capture neighborhood relationships among items. LETTER \cite{LETTER} leverages semantic, collaborative and diversity regularization. The first two types of regularization can extract semantic information and collaborative signals, respectively. The diversity regularization can mitigate code assignment bias. To consider the multi-modal features of items, MMGRec \cite{MMGRec} leverages a hierarchical quantization method (i.e., Graph RQ-VAE) by incorporating multi-modal features. SpecGR \cite{SpecGR} leverages a drafter model to generate some candidate items, including some existing items and new items. Meanwhile, SpecGR \cite{SpecGR} leverages a verifier to decide whether to accept or reject candidate items, enabling SpecGR to recommend new items. Compared with traditional SR models, generative recommender models have some advantages as follows:
	\begin{itemize}
		\item Generative recommender models have strong capabilities in recommending unpopular items. Therefore, they can alleviate the cold-start problem.
		\item By leveraging temperature-based sampling, generative recommender models can generate diverse recommendation lists.
	\end{itemize}
	However, generative recommender models also have some disadvantages. For example, they might generate invalid semantic IDs during the prediction phase.
	\begin{figure*}[h]
		\includegraphics[width=\linewidth]{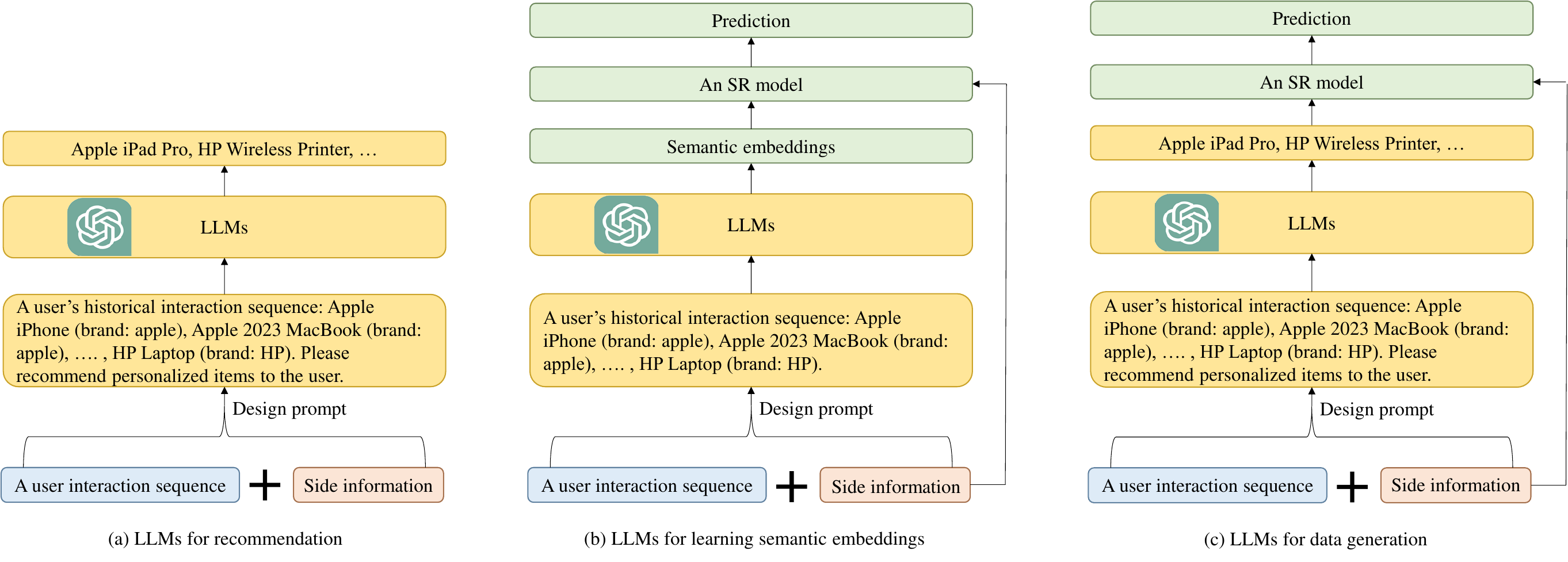}
		\caption{The common SR models based on LLMs.}
		\label{fig:LLMs-usage}
	\end{figure*}
	\subsection{LLM-powered SR}
	\subsubsection{Large Language Models}
	Large language models (LLMs) can handle more complex tasks compared with traditional language models based on deep learning. The main characteristics of LLMs are their billions of parameters as well as the vast amounts of data and computational resources used during training. Therefore, LLMs have a deeper understanding of natural language and extensive knowledge extracted from amounts of data. There are many kinds of LLMs. In our survey, we introduce them briefly as follows:
	\begin{itemize}
		\item LLaMA: LLaMA \footnote{\url{https://huggingface.co/meta-llama/Llama-2-13b}} (Large Language Model Meta AI) is developed by Meta. It is a family of large language models, including LLaMa1, LLaMa2 and LLaMa3.
		\item ChatGLM: ChatGLM (Generative Language Model) is developed by Tsinghua University, which is a bilingual model for understanding and generating both Chinese and English.
		\item T5: T5 (Text-To-Text Transfer Transformer) is developed by Google, which can solve many kinds of tasks by modeling each task as a text-to-text task.
		\item ChatGPT: ChatGPT \footnote{\url{https://platform.openai.com/}} is the most popular large language model, which is developed by OpenAI. It includes models such as GPT-3.5, GPT-4, GPT-4o, and other variants.
		\item PaLM: PaLM (Pathways Language Model) is developed by Google, which is one of the largest language models. It shows strong capabilities in multilingual tasks and source code generation.
		\item Vicuna: Vicuna \footnote{\url{https://github.com/lm-sys/FastChat}} is designed by fine-tuning LLaMA on high-quality conversational data.
		\item Qwen: Qwen is developed by Alibaba and excels at processing some tasks described in both Chinese and English.
		\item Baichuan: Baichuan contains a series of large language models developed by Baichuan Intelligent Technology, including Baichuan 2 and Baichuan-M1.
		\item DeepSeek: DeepSeek \footnote{\url{https://github.com/deepseek-ai}} has recently achieved remarkably impressive performance on various metrics in comparison with the most powerful LLMs.
	\end{itemize}
	Large language models have a strong ability to understand text descriptions and integrate external knowledge. Consequently, incorporating LLMs into sequential recommendation (SR) models can enhance performance and alleviate challenges such as cold-start and data sparsity problems. As shown in Table~\ref{tab:SR_methods_LLMs}, there are some SR models leveraging LLMs for recommendation. The specific details are shown as follows:
	\begin{table*}[h]
		\caption{The SR models based on LLMs.}
		\label{tab:SR_methods_LLMs}
		\centering
		\begin{tabular}{p{5cm}l}
			\hline
			Large Language Models & Representative Works\\
			\hline
			T5  & \cite{P5}, \cite{Lite-LLM4Rec}, \cite{POD}\\
			LLaMA & \cite{SLIM}, \cite{LLM-TRSR}, \cite{RecInterpreter}, \cite{TALLRec}, \cite{DLLM2Rec}, \cite{E4SRec}, \cite{URLLM}, \cite{SeRALM}, \cite{SAID}, \cite{GenRec}, \cite{LLM4DSR}, \cite{SLMRec}, \cite{LLMEmb}\\
			ChatGLM & \cite{LSVCR}\\
			ChatGPT & \cite{LLM-ESR}, \cite{FineRec}, \cite{LRD}\\
			Vicuna & \cite{ReLLa}, \cite{RecLoRA}\\
			Qwen & \cite{BAHE}\\
			Baichuan & \cite{LLMCDSR}\\
			\hline
		\end{tabular}
	\end{table*}
	\subsubsection{LLMs for Recommendation}
	There are several models leveraging LLMs for recommendation. Some specific details are shown in Figure~\ref{fig:LLMs-usage}. The first type of models leverages LLMs to recommend items to users directly, which is formalized below:
	\begin{equation}
		\hat{x}_{prompt} = \operatorname{LLMs}(x_{prompt})
	\end{equation}
	By reading the outputs of LLMs $\hat{x}_{prompt}$, we can identify the names of recommended items.
	
	The second type utilizes semantic embeddings extracted from LLMs to recommend personalized items.  As shown in part (b) of Figure~\ref{fig:LLMs-usage}, the prompt $x_{prompt}$ might contain the text descriptions about a user's whole historical interaction sequence. Similar to the modal-based SR models, the prompt $x_{prompt}$ might be the text descriptions of item $i$, and LLMs are leveraged to extract semantic information about item $i$, which is formalized as follows:
	\begin{equation}
		\mathbf{e}_{i} = \operatorname{LLMs}(x_{prompt})
	\end{equation}
	Compared with conventional text feature extractors such as BERT, LLMs can extract richer semantic information.
	
	As shown in part (c) of Figure~\ref{fig:LLMs-usage}, the third type of models leverages LLMs to generate high-quality data which is then combined with the original data to train SR models. The data generation process is formalized as follows:
	\begin{equation}
		\hat{x}_{prompt} = \operatorname{LLMs}(x_{prompt})
	\end{equation}
	Compared with the above-mentioned first type, the outputs of LLMs are regarded as augmented data for training. It is noted that the $x_{prompt}$ might vary with the usage of LLMs changing.
	
	The most straightforward approach is to directly leverage LLMs for recommending personalized items. Given user $u$'s historical sequence and a target item, LLMs predict $u$'s preferences for the target item. Meanwhile, given the interacted item sequences, LLMs can generate corresponding recommendation lists. For example, P5 \cite{P5} can generate personalized recommend items by inputting historical interaction items into LLMs. There are different tasks, such as generating an item or a list of items, rating items, and ranking the order of the interacted items. POD\cite{POD} converts discrete prompt into a set of continuous prompt vectors. This method can reduce the inference time by bridging the gap between IDs and words. GenRec \cite{GenRec} leverages an adaptation method (LoRA) to fine-tune and perform inference tasks on LLaMA. However, these models have some weaknesses, such as the potential to generate non-existent items.
	\subsubsection{LLMs for Learning Semantic Embeddings}
	Some SR models use LLMs to initialize item embeddings. Firstly, RecInterpreter \cite{RecInterpreter} leverages a linear layer to map item embeddings from an item ID embedding space to a text embedding space. By providing task-specific textual prompts and projected hidden representations, LLaMA is encouraged to generate textual descriptions of items in interaction sequences. The second task aims to discover residual items by task-specific textual prompts and hidden representations. To solve the problem of long-tail items, LLM-ESR \cite{LLM-ESR} leverages LLMs to generate semantic embeddings for enhancing the semantic information of items. Common methods are to combine item ID embeddings and generated semantic embeddings for achieving better recommendation performance. To combine the advantages of traditional SR models and LLMs, E4SRec \cite{E4SRec} uses item IDs as input. This method ensures that all the generated and recommended items fall within candidate lists. Therefore, E4SRec can handle item IDs more effectively. LLMEmb \cite{LLMEmb} is a novel method that leverages LLMs to generate item embeddings. To integrate collaborative signals into LLM-generated embeddings, LLMEmb leverages a recommendation adaptation training strategy. Though knowledge can be transferred from some source domains to a target domain, the cold-start problem cannot be completely solved. LLMs cannot capture collaborative information effectively. Traditional cross-domain sequential recommendation models can capture collaborative signals while LLMs-based models excel at capturing semantic information. URLLM \cite{URLLM} incorporates both advantages. Firstly, URLLM leverages a dual-graph neural network to capture collaborative signals and transfer knowledge from a source domain to a target domain. Meanwhile, URLLM leverages a retrieve-generation model to incorporate structure information into an LLMs-based recommender model. Finally, URLLM combines collaborative information and semantic information to recommend items. LLMCDSR \cite{LLMCDSR} leverages LLMs to predict unobserved cross-domain interactions, which improves the recommendation performance for non-overlapped users. LLMCDSR \cite{LLMCDSR} leverages a collaborative-textual contrastive pre-training approach to integrate collaborative information into textual features. From these models, we can draw the following conclusions: 1) Initializing item embeddings with a large language model can obtain recommendation performance gains. 2) Fine-tuning LLMs for recommendation tasks enables LLMs to learn domain-specific and domain-shared knowledge.
	
	To reduce computational overload and enhance the efficiency and effectiveness of inference, Lite-LLM4Rec \cite{Lite-LLM4Rec} leverages a beam search decoding method and a hierarchical LLM structure for recommendation. We can leverage prompts to conduct fine-tuning of LLMs. TALLRec \cite{TALLRec} employs two tuning methods, such as Alpaca tuning and rec-tuning, to optimize the tuning process of LLMs more efficiently. SeRALM \cite{SeRALM} leverages item IDs and knowledge generated by LLMs as input. Meanwhile, SeRALM leverages the feedback from recommender systems to refine the knowledge of items generated by LLMs. To extract semantic information generated by LLMs and enhance computational efficiency, SAID \cite{SAID} leverages a projector to map item IDs into item embeddings and feeds them into LLMs for refining the item embeddings. Finally, SAID leverages the trained projector to recommend items more effectively.
	
	Knowledge distillation and contrastive learning are effective techniques for extracting knowledge from LLMs. Knowledge distillation can enhance the efficiency of LLMs. SLIM \cite{SLIM} can obtain semantic knowledge from LLMs through knowledge distillation. DLLM2Rec \cite{DLLM2Rec} distills the knowledge from LLMs-based SR models to conventional SR models by leveraging two kinds of distillation (i.e., collaborative embedding distillation and importance-aware ranking distillation). LSVCR \cite{LSVCR} is trained in two stages (i.e., personalized preferences alignment and recommendation-oriented fine-tuning). It uses contrastive learning between preferences generated by LLMs-based recommendation models and preferences generated by conventional SR models. Therefore, these conventional SR models can obtain richer semantic information. SLMRec\cite{SLMRec} leverages a layer-wise knowledge distillation approach to enhance the efficiency of SR models based on LLMs.
	
	LLMs often struggle with processing long text descriptions. LLM-TRSR \cite{LLM-TRSR} can leverage too-long text descriptions as the input of LLMs by segmenting original users' interaction sequences. The LLMs-based SR models will consume lots of computational resources when processing long interacted sequences. To solve this problem, BAHE \cite{BAHE} leverages a behavior-aggregated hierarchical encoding method to enhance computational efficiency. By encoding a user’s separate interacted sequences in parallel, BAHE can obtain the user’s final representations by concatenating them together. Because LoRA’s shared parameters across all users limit its ability to recommend personalized items. Meanwhile, ultra-long interaction sequences might affect the efficiency of LLMs. Therefore, RecLoRA \cite{RecLoRA}  proposes a personalized LoRA module to overcome this limitation and designs a long-short modality retriever to retrieve interaction sequences of different lengths based on different modalities.
	
	\subsubsection{LLMs for Data Generation}
	LLMs can be leveraged to generate data for alleviating the data sparsity and cold-start problems. For example, KDA$_{LRD}$ \cite{LRD} leverages LLMs to describe relationships among items. Most existing models rely on predefined relations among items, which might result in sparse issues. To generate more item-to-item relations, KDA$_{LRD}$ leverages LLMs to learn item representations. An item reconstruction module leverages the estimated relation and another item to reconstruct the target item. Finally, KDA$_{LRD}$ leverages the predefined and generated relations for recommendation. FineRec \cite{FineRec} initially leverages LLMs to extract attribute pairs from reviews. Then by constructing an attribute-aware graph and aggregating, FineRec can obtain improved user and item embeddings for recommendation. For few-shot recommendation, ReLLa \cite{ReLLa} can be used as a data augmentation method to augment data. ReLLa \cite{ReLLa} leverages LLMs to retrieve the top-$k$ semantically relevant behaviors according to a target item. Leveraging the semantic knowledge of LLMs, ReLLa enhances the quality of data. Finally, ReLLa is trained on a mixed data containing the original data and the retrieval-enhanced data. LLM4DSR \cite{LLM-TRSR} leverages LLMs to identify noisy items and replace them with suggested ones.
	
	\begin{figure}[h]
		\includegraphics[width=\linewidth]{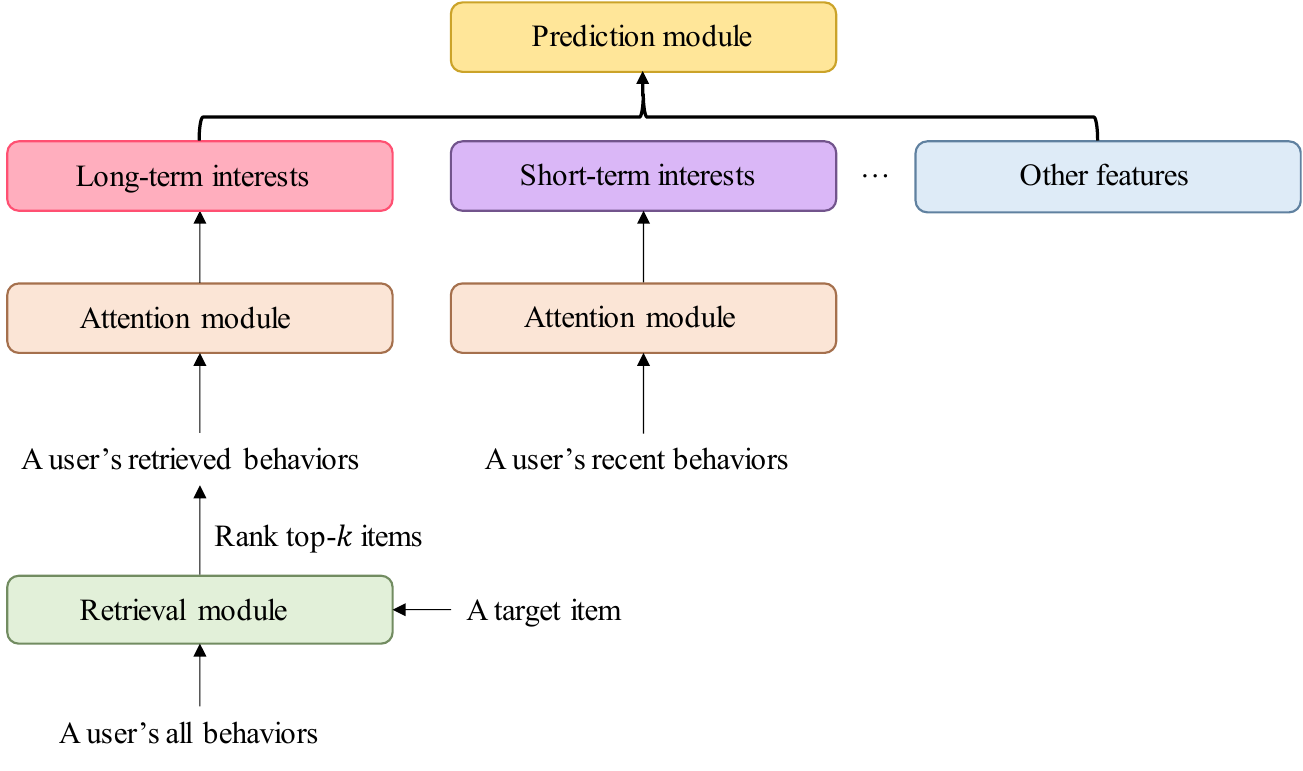}
		\caption{Illustration of a typical ultra-long sequential recommendation method.}
		\label{fig:long-sequence}
	\end{figure}
	\subsection{Ultra-Long SR}
	An ultra-long behavior sequence typically refers to a scenario where the number of interacted items exceeds 1000. Compared with conventional behavior sequence modeling in SR models, ultra-long behavior sequence modeling presents several challenges. Firstly, modeling ultra-long sequences requires substantial computational resources. Secondly, ultra-long sequences often contain noise due to irrelevant behaviors. Finally, during the inference phase, modeling ultra-long sequences can be computationally expensive. For example, the time complexity of Transformer is  $O(L^{2})$. Therefore, SR models while modeling ultra-long sequences might face high latency. The general ultra-long sequence modeling methods are shown in Figure~\ref{fig:long-sequence}. Firstly, the most relevant $k$ items are retrieved from a whole interaction sequence. Then long-term interests are extracted from the retrieved $k$ items by an attention module. Short-term interests are extracted from recently interacted items by an attention module. The process of extracting the two interests is formalized as follows:
	\begin{equation}
		\mathbf{P}_{u}^{short} = \operatorname{Att}(S_{u,recent})
	\end{equation}
	\begin{equation}
		\mathbf{P}_{u}^{long} = \operatorname{Att}(\operatorname{Ret}(S_{u}))
	\end{equation}
	The sequences fed into the attention module are relatively short, so computational efficiency can be enhanced to some extent. Finally, by leveraging the long-term interests, the short-term interests and other features, SR models can recommend personalized items for users.
	
	Traditional SR methods focus on leveraging the most recent behaviors, while UBR4CTR \cite{UBR4CTR} emphasizes the most relevant behaviors. When only the recent interactions are leveraged, the long-term dependencies and periodicity might not be captured accurately. ETA-Net \cite{ETA-Net} aims to retrieve relevant items from long sequences in a more efficient manner. Instead of using a traditional dot-product operation, ETA-Net leverages a hashing-based target attention algorithm to calculate the Hamming distance, and then selects the top-$k$ relevant behavior items for a target item. Finally, ETA-Net leverages multi-head attention to learn long sequence embeddings for capturing diverse interests from the selected top-$k$ items. During prediction, ETA-Net leverages five different embeddings (i.e., long sequence embeddings, target item embeddings, short-sequence embeddings, user profile embeddings and context embeddings). For users with long interaction sequences, these sequences can be divided into sessions based on time intervals between interactions. As a result, HNUM \cite{HNUM} is designed with a two-layer architecture. The first layer is a session-level GRU model designed to capture users’ short-term preferences within the current session. The second layer is a user-level memory network, which can capture users’ long-term preferences across their entire interaction history. SIM \cite{SIM} captures users' interests using both general search unit (GSU) and exact search unit (ESU). GSU leverages both hard search and soft search methods. The hard search aims to retrieve interacted items belonging to the same category of a candidate item from a user’s long sequence. The soft search leverages inner product to identify top-$k$ relevant items. These top-$k$ items are used to construct users’ sub-sequences. By employing multi-head attention, ESU effectively models the relationship between candidate items and the sub-sequences to extract users' long-term preferences. To solve the inconsistency between CP-GSU and ESU for target items, TWIN \cite{TWIN} leverages the same target-behavior relevance metric in both modules. Furthermore, to enhance efficiency, TWIN compresses user-item cross features into bias terms for target attention. To tackle ultra-long user behavior sequences, TWIN-V2 \cite{TWIN-V2} leverages a hierarchical method to compress life-cycle behaviors. Similar items are classified into the same cluster. Meanwhile, TWIN-V2 leverages cluster-aware target attention to learn users’ diverse interests. R-RNN \cite{R-RNN} learns a user’s global interests and  recent interests based on global user behaviors and recent user behaviors, respectively. To improve efficiency while modeling long behavior sequences, SAM \cite{SAM} leverages dual-query attention by regarding target item embedding and memory vector as queries. User interest center (UIC) is designed to decouple user interest modeling from the whole model. Modeling user interests needs to consume lots of computational time. UIC learns a user’s latest interest representations according to real-time user behavior events. MIMN \cite{MIMN}  leverages memory utilization regularization and memory induction unit to learn multi-channel user interests. UBCS \cite{UBCS} proposes a behavior sampling module to sample short sequences based on candidate items, relevance and temporal information. To speed up the process of sampling, UBCS leverages an item clustering module to cluster candidate items for selecting some representative candidate items. CoFARS \cite{CoFARS} takes into account the influence of contexts on user preferences. CoFARS leverages a prototype-based approach to identify contexts that reflect similar user preferences. Then, CoFARS constructs a temporal graph containing context nodes and prototype nodes for integrating temporal information into users’ interests. Finally, CoFARS identifies the prototypes that are consistent with the target context to generate a sub-sequence. Different from single-domain long sequence modeling, LCN \cite{LCN} can transfer knowledge from a source domain to a target domain. In a cross representation production module, LCN leverages contrastive learning to bridge the connections among similar items in different domains. In a lifelong attention pyramid module, LCN leverages three kinds of cascading attentions to extract preferences from user lifelong behavior sequences considering candidate items. LRURec \cite{LRURec} leverages linear recurrent units and a recursive parallelization framework to enhance computational efficiency and realize low-cost inference. LinRec \cite{LinRec} reduces the quadratic time complexity of self-attention to a linear time complexity while preserving the advantages of attention mechanisms. To address the issue of sparse search behaviors failing to capture users’ interests, QIN \cite{QIN} adopts a two-stage search method. Firstly, the top-$k$ items from a long sequence for user $u$ are retrieved using a pre-trained retrieval model. Secondly, QIN predicts whether user $u$ will purchase items based on the retrieved top-$k$ items. For QIN, it first searches sub-sequences based on the relevance between behaviors and queries. Secondly, QIN searches sub-sequences according to the relevance between behaviors and target items. By considering item ID field and attribute field separately, QIN leverages a fused attention unit to learn users’ representations better. To efficiently retrieve sub-sequences, SDIM \cite{SDIM} leverages an effective sampling method to generate sub-sequences. SDIM generates hash signatures for both the candidate item and each item in a user behavior sequence by multiple hash functions. SDIM can obtain the sub-sequences where items have the hash signatures the same as the candidate item. Item IDs might not exist in a data and different modal (i.e., text, images and attributes) embeddings are not aligned. Meanwhile, users’ interacted item sequence and users’ search query sequence reflect users’ interests from different perspectives. To address these problems, SEMINAR \cite{SEMINAR} leverages a pre-training search unit to learn query-item pairs. Then, SEMINAR leverages pre-trained embeddings to initialize the network. Additionally, SEMINAR leverages a multi-modal product quantization strategy to reduce time complexity. The basic framework of SEMINAR includes GSU and ESU. To model users’ long sequences and capture evolving users’ interests, HPMN \cite{HPMN}  leverages memory slots to realize sequential patterns’ personalized memorization for each user. To capture multi-scale and evolving sequence patterns, HPMN leverages a hierarchical incremental updating mechanism.
	\subsection{Data-Augmented SR}
	There are two methods to conduct data augmentation. The first augmented methods only consider the information about item IDs. The second augmented methods that consider the role of side information. These approaches aim to alleviate the data sparsity problem and learn high-quality user representations. There are some SR models using data-augmented methods as follows.
	
	Time intervals in users' interaction sequences might vary significantly, which might fail to model users’ preferences due to preference drift. To solve this problem, TiCoSeRec \cite{TiCoSeRec} standardizes interaction sequences by transforming uneven time intervals into uniform distributions. TiCoSeRec consists of five operators (i.e., Ti-Crop, Ti-Reorder, Ti-Mask, Ti-Substitute and Ti-Insert). Different from traditional operators, the five data augmentation methods explicitly account for time intervals. Finally, TiCoSeRec leverages contrastive learning to ensure a high similarity between the augmented and original sequences. CL4SRec \cite{CL4SRec} leverages three kinds of data augmentation approaches (i.e., item crop, item mask and item reorder). Meanwhile, CL4SRec leverages contrastive learning between one augmented user interaction sequence and another augmented one. To alleviate the data sparsity problem and obtain stable users’ interests, TGCL4SR \cite{TGCL4SR} leverages two methods (i.e., neighbor sampling and time disturbance) to obtain augmented subgraphs. The neighbor sampling constructs augmented subgraphs by randomly sampling the neighbor nodes of each item. The time disturbance constructs the hypergraph to randomly add some noise that obeys Gaussian distribution to timestamps. Finally, by using disturbed graph contrastive loss and subgraph contrastive loss, high-quality representations can be obtained. Driven by the potential of data-centric AI, DR4SR+ \cite{DR4SR+} incorporates a model-aware data personalizer to tailor the regenerated data. In the process of model-agnostic data regeneration, a regenerator is pre-trained by learning transition patterns obtained through rule-based methods in the original interaction sequences. DR4SR+ leverages a diversity-promoted regenerator to capture one-to-many mapping relationships between the original sequences and generated patterns. To generate a personalized data for target items, DR4SR+ uses a data personalizer to evaluate the quality of each generated sample for a target model. Traditional data augmentation methods often modify an original interaction sequence randomly. For example, some items are masked in sequences randomly. These methods might break sequential correlations. To solve this problem, MStein \cite{MStein} leverages Wasserstein discrepancy measurement to measure mutual information among augmented sequences and ensures the maximum of the mutual information. MBASR \cite{MBASR} considers users’ behaviors while data augmentation is conducted. DuoRec \cite{DuoRec} leverages a contrastive learning regularizer to generate item embeddings with more uniform magnitude and frequency. END4Rec \cite{END4Rec} improves the performance of Transformer by pruning noisy input.  ASReP \cite{ASReP} leverages a pre-training Transformer to enhance the length of short  users' interaction sequences, and then the Transformer is fine-tuned by augmented sequences.
	\section{Empirical Studies}\label{sec:empirical_studies}
	In this section, our survey will introduce some datasets, evaluation protocols and evaluation metrics used in sequential recommendation. Finally, we will present experimental results of some representative SR models.
	\subsection{Datasets}
	Training modality-based SR models with LLMs requires datasets that include multi-modal features. In our survey, we introduce some public datasets containing multi-modal features. Some specific details are shown in Table~\ref{tab:datasets_mm_f}.

\begin{table}[h]
	\caption{The statistics about several widely used datasets.}
	\label{tab:datasets_mm_f}
	\centering
	\begin{tabular}{ll}
		\hline
		Datasets & Multi-modal Features\\
		\hline
		MIND\tablefootnote{\url{https://msnews.github.io}} & text \\
		H-M\tablefootnote{\url{https://www.kaggle.com/c/h-and-m-personalized-fashion-recommendations/data}} & text, images \\
		Bili\tablefootnote{\url{https://github.com/westlake-repl/NineRec?tab=readme-ov-file}} & text, images \\
		Art of the Mix\tablefootnote{\url{https://brianmcfee.net/data/aotm2011.html}} & text \\
		LastFM\tablefootnote{\url{https://www.kaggle.com/datasets/harshal19t/lastfm-dataset}} & text \\
		Amazon 2023\tablefootnote{\url{https://huggingface.co/datasets/McAuley-Lab/Amazon-Reviews-2023}} & text, images, videos \\
		MicroLens\tablefootnote{\url{https://github.com/westlake-repl/MicroLens}} & text, images, videos \\
		MovieLens\tablefootnote{\url{https://grouplens.org/datasets/movielens/}} & text \\
		Steam\tablefootnote{\url{https://huggingface.co/datasets/FronkonGames/steam-games-dataset}} & text, images \\
		Yelp\tablefootnote{\url{https://www.yelp.com/dataset}} & text, images \\
		Goodreads\tablefootnote{\url{https://www.kaggle.com/datasets/jealousleopard/goodreadsbooks}} & text, images \\
		Douban\tablefootnote{\url{https://www.kaggle.com/datasets/fengzhujoey/douban-datasetratingreviewside-information}} & text \\
		\hline
	\end{tabular}
\end{table}

% \begin{table*}[h]
% 	\caption{The statistics about several widely used datasets.}
% 	\label{tab:datasets_mm_f}
% 	\centering
% 	\begin{tabular}{ll}
% 		\hline
% 		Datasets & Multi-modal Features \\
% 		\hline
% 		MIND\footnotemark[1] & text \\
% 		H-M\footnotemark[2] & text, images \\
% 		Bili\footnotemark[3] & text, images \\
% 		Art of the Mix\footnotemark[4] & text \\
% 		LastFM\footnotemark[5] & text \\
% 		Amazon 2023\footnotemark[6] & text, images, videos \\
% 		MicroLens\footnotemark[7] & text, images, videos \\
% 		MovieLens\footnotemark[8] & text \\
% 		Steam\footnotemark[9] & text, images \\
% 		Yelp\footnotemark[10] & text, images \\
% 		Goodreads\footnotemark[11] & text, images \\
% 		Douban\footnotemark[12] & text \\
% 		\hline
% 	\end{tabular}
% \end{table*}
% 
% % 脚注内容
% \footnotetext[1]{\url{https://msnews.github.io}}
% \footnotetext[2]{\url{https://www.kaggle.com/c/h-and-m-personalized-fashion-recommendations/data}}
% \footnotetext[3]{\url{https://github.com/westlake-repl/NineRec?tab=readme-ov-file}}
% \footnotetext[4]{\url{https://brianmcfee.net/data/aotm2011.html}}
% \footnotetext[5]{\url{https://www.kaggle.com/datasets/harshal19t/lastfm-dataset}}
% \footnotetext[6]{\url{https://huggingface.co/datasets/McAuley-Lab/Amazon-Reviews-2023}}
% \footnotetext[7]{\url{https://github.com/westlake-repl/MicroLens}}
% \footnotetext[8]{\url{https://grouplens.org/datasets/movielens/}}
% \footnotetext[9]{\url{https://huggingface.co/datasets/FronkonGames/steam-games-dataset}}
% \footnotetext[10]{\url{https://www.yelp.com/dataset}}
% \footnotetext[11]{\url{https://www.kaggle.com/datasets/jealousleopard/goodreadsbooks}}
% \footnotetext[12]{\url{https://www.kaggle.com/datasets/fengzhujoey/douban-datasetratingreviewside-information}}

	These datasets primarily contain three different modals (i.e., text, images and videos). The strong sequential structure of these datasets is important for training SR models. Compared with other datasets in Table~\ref{tab:datasets_mm_f}, Yelp and Steam might have weak sequential structure \cite{sequential_structure}.
	\subsection{Evaluation Protocols}
	Datasets can be divided into training, validation and testing data using two primary methods. An early method is the leave-one-out strategy \cite{SASRec}. For this split method, all users' last interaction is split into the testing data, all users' penultimate interaction is split into the validation data, and the remaining interactions are split into the training data. This split method is simple and efficient, but it might lead to the problem of data leakage \cite{split_by_timestamp}. This is because interactions in the validation data may precede those in the training data based on their timestamps. The second method obtains the training, validation and testing data by specific timestamps \cite{CCA}. We can set two timestamps (i.e., $T1$ and $T2$). All interactions happening before timestamp $T1$ are split into the training data. All interactions happening between timestamp $T1$ and timestamp $T2$ are split into the validation data. All interactions happening after timestamp $T2$ are assigned to the testing data. This splitting method helps to prevent data leakage. However, this splitting method has a shortcoming that some users might not be evaluated. Because all interacted records for some users could fall within training data.
	
	During the validation and testing phase, candidate items can be selected by three methods (i.e., random sampling, sampling based on popularity, and full sampling). Random sampling selects a small number of candidate items randomly \cite{SASRec}. This method is simple yet effective. Because the SR models are usually deployed in ranking or re-ranking phase, there are only a small number of candidate items. The number of candidate items is consistent with the actual situation. However, random sampling might lead to variability in the candidate items. The evaluation results might be unstable with the varying candidate items. Sampling based on popularity is that the candidate items are selected based on popularity \cite{MGCL}. If the popularity of items is larger, the probability of these items that are selected as candidate items is larger. This method can reduce the influence of randomness. Users interact with popular items easily, therefore these items appear in candidate items more frequently. Compared with random sampling, this split method is more stable and provides a more accurate evaluation of recommendation performance. The final method considers all items in the dataset as candidate items \cite{DR4SR+}. This method can evaluate the recommendation performance of different models from a comprehensive view. However, it might consume lots of computational resources especially when the dataset contains a large number of items.
	\subsection{Evaluation Metrics}
	SR models aim to generate a top-$N$ recommendation list. The evaluation metrics mainly aim to measure whether the user's most recently interacted item is in the top-$N$ recommendation list and the position in the top-$N$ recommendation list. Two commonly used evaluation metrics \cite{SASRec,Boka.al.} are shown as follows.
	
	Hit Ratio (HR) denotes the ratio of ground-truth items included in the top-$N$ recommendation list. It can be formalized as:
	\begin{equation}
		HR@N = \frac{1}{|\mathcal{U}|}\sum_{u \in \mathcal{U}}\sigma(R_{u,t_{u}} \leq N)
	\end{equation}
	where $\sigma(\cdot)$ is an indicator function. If the rank generated by an SR model is among top-$N$, the value of $\sigma(\cdot)$ is 1, otherwise is 0.
	Normalized Discounted Cumulative Gain (NDCG) evaluates the quality of the ranking by considering both the relevance and position of ground-truth items in the top-$N$ recommendation list. It can be formalized as:
	\begin{equation}
		NDCG@N = \frac{1}{\mathcal{|U|}}\sum_{u \in \mathcal{U}}\frac{\sigma(R_{u,t_{u}} \leq N)}{\log_{2}(1 + R_{u,t_{u}})}
	\end{equation}
	\subsection{Experimental Results}
	% Table generated by Excel2LaTeX from sheet 'Sheet3'
	\begin{table*}[htbp]
		\caption{Experimental results of different SR models. The results are copied from papers \cite{MISSRec,SpecGR,Lite-LLM4Rec} for direct comparison, where  “-” means that the corresponding experimental results are not available in the original papers.}
		\centering
		\label{tab:experiment_results}%
		
		\begin{tabular}{ccccccc}
			\hline \multirow{2}{*}{Models} & \multicolumn{2}{c}{Office} & \multicolumn{2}{c}{Game} & \multicolumn{2}{c}{Toy}  \\		
			\cmidrule(lr){2-3} \cmidrule(lr){4-5} \cmidrule(lr){6-7} 	
			& \multicolumn{1}{l}{HR@10} & \multicolumn{1}{l}{NDCG@10} & \multicolumn{1}{l}{HR@10} & \multicolumn{1}{l}{NDCG@10} & \multicolumn{1}{l}{HR@10} & \multicolumn{1}{l}{NDCG@10} \\
			\hline
			SASRec & 0.1056 & 0.0710 & 0.0186 & 0.0093 & 0.0561 & 0.0312 \\
			BERT4Rec & 0.0825 & 0.0634 & -     & -     & 0.0352 & 0.0179 \\
			\hline
			FDSR  & 0.1118 & 0.0868 & 0.0190 & 0.0101 & 0.0568 & 0.0344 \\
			S$^{3}$-Rec & 0.1030 & 0.0653 & 0.0195 & 0.0094 & 0.0538 & 0.0276 \\
			\hline
			UniSRec w/o IDs & 0.1013 & 0.0619 & -     & -     & - & - \\
			UniSRec & 0.1280 & 0.0831 & 0.0225 & 0.0115 & - & - \\
			MISSRec w/o IDs & 0.1258 & 0.0795 & -     & -     & - & - \\
			MISSRec & 0.1301 & 0.0842 & -     & -     & - & - \\
			RECFORMER & -     &  -    & 0.0243 & 0.0114 & - & - \\
			\hline
			TIGER & -     &  -    & 0.0222 & 0.0114 & - & - \\
			SpecGR & -     &  -    & 0.0229 & 0.0115 & - & - \\
			\hline
			P5    & -     &  -    & -     & -     & 0.0543 & 0.0356 \\
			POD   & -     &  -    & -     & -     & 0.0566 & 0.0421 \\
			Lite-LLM4Rec & -     &  -    & -     & -     & 0.0682 & 0.0488 \\
			\hline
		\end{tabular}%
	\end{table*}
	
To show the relative recommendation performance of different SR models, we quote and copy the results of some recent works \cite{MISSRec,SpecGR,Lite-LLM4Rec}. The compared SR models can be classified into five categories, including pure ID-based SR (i.e., SASRec \cite{SASRec} and BERT4Rec \cite{BERT4Rec}), SR with side information (i.e., FDSA \cite{FDSA} and S$^{3}$Rec \cite{S3-Rec}), multi-modal SR (i.e., UniSRec \cite{UniSRec}, MISSRec \cite{MISSRec} and RECFORMER \cite{RECFORMER}), generative SR (i.e., TIGER \cite{TIGER} and SpecGR \cite{SpecGR}), and LLM-powered SR (i.e., P5 \cite{P5}, POD \cite{POD} and Lite-LLM4Rec \cite{Lite-LLM4Rec}). ``Office", ``Game" and ``Toy" are three domains in an Amazon dataset. Two widely-used metrics (i.e., HR@10 and NDCG@10) are adopted for performance evaluation. In each domain, the leave-one-out strategy is leveraged to obtain the training, validation and test data. Meanwhile, all items are taken as the candidate ones in recommendation. The data in each domain is processed according to the corresponding papers \cite{MISSRec,SpecGR,Lite-LLM4Rec}. From the experimental results in Table~\ref{tab:experiment_results}, we can have the following observations:
	\begin{itemize}
		\item SR models with side information achieve better performance than pure ID-based SR ones, which demonstrates that leveraging item features can alleviate the cold-start and sparsity problem to some extent.
		\item By removing item IDs from multi-modal SR models, the recommendation performance will decrease. It is because that the item IDs can capture the collaborative signals and users' fine-grained preferences.
		\item LLM-powered SR models show better recommendation performance than traditional SR models. We can see that the semantic knowledge extracted by LLMs are useful for enhancing the recommendation performance.
		\item Compared with pure ID-based and multi-modal SR models, generative SR models show comparable and even better performance. By leveraging semantic IDs, generative models can capture different grained semantic information.
	\end{itemize}
	\section{Prospects and Future Directions}\label{sec:prospects_and_future_directions}
	SR models have been studied for several years, but there are certain areas that remain to be fully explored.
	\subsection{Open-Domain SR}
	Traditional SR models are trained on a single and static data with predefined features. When new items emerge frequently and data resources are increasingly diverse, the traditional SR models are difficult to recommend personalized items. Leveraging the multi-modal features of items and LLMs, open-domain SR models can operate effectively in dynamic environments with a continuous influx of new items. Meanwhile, the open-domain SR models can be transferred across different platforms and domains. By transferring knowledge from external domains or platforms, the open-domain SR models can achieve better recommendation performance within the target domain or platform.
	\subsection{Data-Centric SR}
	To improve the recommendation performance, data-centric SR models mainly focus on improving the quality of data rather than refining model structures. The data is important in sequential recommendation tasks. The data-centric SR models mainly leverage three techniques (i.e., data generation, data denoising and data debiasing). The data generation has been researched for many years.  However, challenges remain in generating high-quality data for certain existing datasets. The quality of the generated data cannot be measured by a specific method. Some SR models leverage data augmentation to generate additional data, aiming to improve the recommendation performance. However, SR models need to be trained in a larger number of data, which might consume more computational resources and training time. Therefore, we need to design a data augmentation method to improve the quality of generated data and reduce the training time. In the meantime, we can consider some multi-modal features to generate some high-quality data.
	\subsection{Cloud-Edge Collaborative SR}
	Cloud-edge collaborative SR models combine the strengths of cloud-based and edge-based SR models. Compared with conventional SR models, cloud-edge collaborative SR models not only can improve the recommendation performance but also protect user privacy and leverage resources more effectively. The conventional SR models often require lots of computational resources. Therefore, the conventional SR models are hard to deploy on their own computers. Moreover, transmitting interaction data to a server could risk exposing private information. Therefore, the cloud-edge collaborative SR models are meaningful. Federated SR models are one such approach specifically designed to protect privacy.
	\subsection{Continuous SR}
	In many short-video scenarios, such as TikTok, the length of a user's interaction sequence tends to grow over time. Some ultra-long interaction sequences might influence the efficiency of SR models. Therefore, how to design an SR model that can handle ultra-long interaction sequences is a challenge. Meanwhile, a user's interests might evolve over time. Therefore, continuous SR models need to update the user's interests dynamically based on recent interactions.
	\subsection{SR for Good}
	Some existing SR models aim to improve economic growth. However, SR models for good aim to recommend personalized items benefit both individuals and society while also supporting economic development. For example, to reduce carbon emissions, it is important to recommend environmentally friendly items for users. In educational field, SR models should recommend learning materials to help users learn more effectively. In healthy field, SR models should recommend personalized exercises to help users improve their physical fitness and overall health. The SR models for good remain to be explored.
	\subsection{Explainable SR}
	Some SR models based on multi-modal features and large language models focus on improving the recommendation accuracy but often fail to provide explanations for why these related items are recommended. LLMs excel at processing and understanding multi-modal features (i.e., text, images and so on), which provides an opportunity to generate some explanations. Therefore, in future work, we can leverage large language models and multi-modal features of interacted items to offer more interpretation in recommendations.
	\section{Conclusions}\label{sec:conclusions}
	In this survey, we have shown that SR models have achieved significant advancements in recommending personalized items for users. From their development, we can have some observations. Firstly, SR models combining IDs and general features can tackle the data sparsity and cold-start problems better than pure ID-based SR models. Secondly, modal-based SR models are often associated with better transferability than pure ID-based SR models and SR models combining IDs and general features. Thirdly, modal-based SR models are of high time complexity and need to consume more training time than pure ID-based SR models. Fourthly, multi-modal SR models using IDs can learn fine-grained preferences and achieve more accurate recommendation performance. Fifthly, LLMs-based SR models and generative SR models show further potential for improving recommendation effectiveness. Sixthly, SR models can effectively tackle ultra-long interaction sequences by retrieving a few items from these sequences. Finally, data augmentation methods can alleviate the data sparsity and cold-start problems to some extent.
	
	As for future directions, we discuss several interesting topics worthy of exploration, which are about scenario, data, architecture, time, society and explainability.
	\begin{acknowledgement}
	We thank the support of National Natural Science Foundation of China (Nos. 62461160311, 62172283 and 62272315), and Guangdong Basic and Applied Basic Research Foundation (Grant No. 2024A1515010122).
	\end{acknowledgement}
	\bibliographystyle{fcs}
	\bibliography{no_series_ref}
	\begin{biography}{LiweiPan}
		Liwei Pan received the BS degree from the Hubei University of Technology, China, in 2023. He is currently working toward the Ph.D. degree with the College of Computer Science and Software Engineering, Shenzhen University, Shenzhen, China. His research interests include recommender systems, deep learning, and transfer learning. 
	\end{biography}
	\begin{biography}{WeikePAN}
		Weike Pan received the Ph.D. degree in Computer Science and Engineering from the Hong Kong University of Science and Technology, Kowloon, Hong Kong, China, in 2012. He is currently a professor with the College of Computer Science and Software Engineering, Shenzhen University, Shenzhen, China. His research interests include recommender systems, deep learning, transfer learning and federated Learning. 
	\end{biography}
	\begin{biography}{MeiyanWei}
		Meiyan Wei will join the College of Computer Science and Software Engineering of Shenzhen University as a Master student in Fall 2025.  Her research interests include recommender systems and deep learning.
	\end{biography}
	\begin{biography}{HongzhiYIN}
		Hongzhi Yin received the PhD degree in Computer Science from Peking University, in 2014. He is a full professor and director of the Responsible Big Data Intelligence Lab (RBDI), University of Queensland. He is an ARC future fellow and an ARC DECRA fellow. His research interests include recommendation system, user profiling, topic models, deep learning, social media mining, and location-based services.
	\end{biography}
	\begin{biography}{ZhongMING}
		Zhong Ming received the Ph.D. degree in Computer Science and Technology from the Sun Yat-Sen University, Guangzhou, China, in 2003. He is currently a professor with the College of Big Data and Internet, Shenzhen Technology University, China. His research interests include software engineering and artificial intelligence. 
	\end{biography}
\end{document}